\documentclass[floatfix,aps,twocolumn,showpacs,superscriptaddress,amsmath,amssymb,longbibliography]{revtex4-1}

\usepackage[autostyle=true]{csquotes}

\usepackage{times}

\usepackage{mathtools}
\usepackage{textcomp}
\usepackage{gensymb}
\usepackage{graphicx}

\usepackage{siunitx}
\usepackage{ulem}
\usepackage{physics}
\usepackage{appendix}

\usepackage[unicode=true,
 bookmarks=true,bookmarksnumbered=true,bookmarksopen=true,bookmarksopenlevel=2,
 breaklinks=false,pdfborder={0 0 1},backref=false,colorlinks=true]
 {hyperref}
 
 \hypersetup{linkcolor=blue, citecolor=blue, urlcolor=blue, filecolor=blue, pdfpagelayout=OneColumn,
  pdfnewwindow=true, pdfstartview=XYZ, plainpages=false}

\usepackage[all]{hypcap}

\usepackage{color}

\makeatletter
\def\mathcolor#1#{\@mathcolor{#1}}%
\def\@mathcolor#1#2#3{%
  \protect\leavevmode%
  \begingroup\color#1{#2}#3\endgroup%
}%
\newcommand{\msout}[1]{\text{\sout{\ensuremath{#1}}}}%

\newcommand{\sam}[2]{%
\ifmmode%
  \msout{#1}\mathcolor{blue}{#2}%
\else%
  \sout{#1}\textcolor{blue}{#2}%
\fi}
\newcommand{\samO}[2]{%
\ifmmode%
  \msout{#1}\mathcolor{magenta}{#2}%
\else%
  \sout{#1}\textcolor{magenta}{#2}%
\fi}

\newcommand{\cyr}[2]{%
\ifmmode%
  \msout{#1}\mathcolor{red}{#2}%
\else%
  \sout{#1}\textcolor{red}{#2}%
\fi}

\newcommand{\gab}[2]{%
\ifmmode%
  \msout{#1}\mathcolor{green}{#2}%
\else%
  \sout{#1}\textcolor{green}{#2}%
\fi}

\makeatother

\makeatletter

\newcommand\Autoref[1]{\@first@ref#1,@}
\def\@throw@dot#1.#2@{#1}
\def\@set@refname#1{
    \edef\@tmp{\getrefbykeydefault{#1}{anchor}{}}%
    \xdef\@tmp{\expandafter\@throw@dot\@tmp.@}%
    \ltx@IfUndefined{\@tmp autorefnameplural}%
         {\def\@refname{\@nameuse{\@tmp autorefname}s}}%
         {\def\@refname{\@nameuse{\@tmp autorefnameplural}}}%
}
\def\@first@ref#1,#2{%
  \ifx#2@\autoref{#1}\let\@nextref\@gobble
  \else%
    \@set@refname{#1}
    \@refname~\ref{#1}
    \let\@nextref\@next@ref
  \fi%
  \@nextref#2%
}
\def\@next@ref#1,#2{%
   \ifx#2@ and~\ref{#1}\let\@nextref\@gobble
   \else, \ref{#1}
   \fi%
   \@nextref#2%
}

\makeatother

\begin{document}

\def\be{\begin{equation}}
\def\ee{\end{equation}}

\def\myvec#1{{\bf #1}}
\def\Esw{\myvec{E}_{sw}}
\def\Hsw{\myvec{H}_{sw}}

\title{Chiral thermodynamics in tailored chiral optical environments}

\author{Gabriel Schnoering}
\altaffiliation[Present address: ]{Laboratory of Thermodynamics in Emerging Technologies, ETH Z\"urich, Sonneggstrasse 3, CH-8092 Z\"urich, Switzerland}
\affiliation{Universit\'e de Strasbourg, CNRS, Institut de Science et d'Ing\'enierie Supramol\'eculaires, UMR 7006, F-67000 Strasbourg, France}
\author{Samuel Albert}
\affiliation{Universit\'e de Strasbourg, CNRS, Institut de Science et d'Ing\'enierie Supramol\'eculaires, UMR 7006, F-67000 Strasbourg, France}
\author{Antoine Canaguier-Durand}
\altaffiliation[Present address: ]{Saint-Gobain Research Paris, 39 quai Lucien Lefranc, F-93300 Aubervilliers, France}
\affiliation{Universit\'e de Strasbourg, CNRS, Institut de Science et d'Ing\'enierie Supramol\'eculaires, UMR 7006, F-67000 Strasbourg, France}
\author{Cyriaque Genet}
\email[]{genet@unistra.fr}
\affiliation{Universit\'e de Strasbourg, CNRS, Institut de Science et d'Ing\'enierie Supramol\'eculaires, UMR 7006, F-67000 Strasbourg, France}

\date{\today}

\begin{abstract}

We present an optomechanical model that describes the stochastic motion of an overdamped chiral nanoparticle diffusing in the optical bistable potential formed in the standing-wave of two counter-propagating Gaussian beams. We show how chiral optical environments can be induced in the standing-wave with no modification of the initial bistability by controlling the polarizations of each beam. Under this control, optical chiral densities and/or an optical chiral fluxes are generated, associated respectively with reactive vs. dissipative chiral optical forces exerted on the diffusing chiral nanoparticle. This optomechanical chiral coupling bias the thermodynamics of the thermal activation of the barrier crossing, in ways that depend on the nanoparticle enantiomer and on the optical field enantiomorph. We show that reactive chiral forces, being conservative, contribute to a global, enantiospecific, change of the Helmholtz free energy bistable landscape. In contrast, when the chiral nanoparticle is immersed in a dissipative chiral environment, the symmetry of the bistable potential is broken by non-conservative chiral optical forces. In this case, the chiral electromagnetic fields continuously transfer, through dissipation, mechanical energy to the chiral nanoparticle. For this chiral nonequilibrium steady-state, the thermodynamic changes of the barrier crossing take the form of heat transferred to the thermal bath and yield chiral deracemization schemes that can be explicitly calculated within the framework of our model. Three-dimensional stochastic simulations confirm and further illustrate the thermodynamic impact of chirality. Our results reveal how chiral degrees of freedom both of the nanoparticle and of the optical fields can be transformed into true thermodynamics control parameters, thereby demonstrating the significance of optomechanical chiral coupling in stochastic thermodynamics.

\end{abstract}

\maketitle

\section{Introduction} 

Recently, optomechanical manifestations of chiral light-matter interactions have been explored in the form of new optical forces that stem from the coupling between a chiral object and a chiral electromagnetic field \cite{canaguier2013mechanical,cameron2014discriminatory,ding2014realization,bliokh2014magnetoelectric}. Because they intertwine the chiral content of the electromagnetic field with the chiral response of the object, the new forces are enantioselective and have led to promising chiral sorting strategies \cite{tkachenko2014optofluidic,cameron2014diffraction,canaguier2014chiralnearfields,canaguier2015chiral,Hayat13190,rukhlenko2016completely,kravets2019optical}. These strategies have obviously a strong applicative potential at the nanoscale, when targeting molecular chiral resolution by optomechanical means \cite{Marichez2019}. At these scales, thermal fluctuations impose a stochastic description of these chiral forces that provides an interesting framework for studying the thermodynamics significance of the chiral coupling. It is the purpose of this work to highlight thermodynamic signatures of chirality and thus to address the fundamental question of chirality in the context of stochastic thermodynamics \cite{SekimotoBook,SeifertRPP2012,CilibertoPRX2017,Bechhoefer2020}.

\begin{figure}[htb!]
  \centering{
    \includegraphics[width=0.8\linewidth]{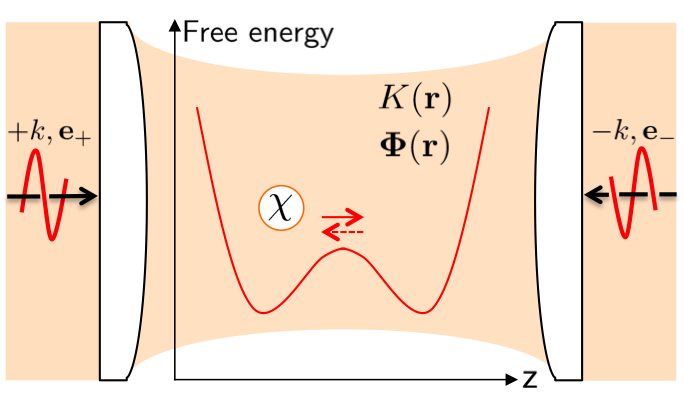}
    }
  \caption{Schematics of the proposed optomechanical model. Two counter-propagating $\pm k$ Gaussian beams, focused to a common waist by two objectives, create a bistable potential free energy surface. A diffusing chiral nanoparticle, optically trapped in this potential in the dipolar regime, is thermally activated and crosses, in both $z\lessgtr 0$ directions, the barrier separating the two potential wells. Depending on the settings of the polarization vectors ${\bf e}_\pm$ of the beams, optical chiral density $K({\bf r})$ and/or chiral flux ${\bf \Phi}({\bf r})$ can be induced in the standing wave between the objectives. When this happens, a chiral coupling involves the chirality of the nanoparticle via the chiral polarizability $\chi$ and the chirality of the field via $K({\bf r})$ and ${\bf \Phi}({\bf r})$, which are respectively time-even pseudoscalar and pseudovector, i.e. truly chiral quantities \cite{BarronLincei}. The chiral coupling generates chiral optical forces that act on the nanoparticle and bias the diffusing motion of the nanoparticle within the bistable potential. This therefore endows the thermally activated barrier crossing with an enantiospecific, chiral discriminative, thermodynamics. }
  \label{fig1a}  
\end{figure}

In a thermodynamic approach, it is interesting to view chiral forces as being induced when a chiral object is immersed within a chiral optical environment. This view indeed draws relevant analogies with chiral chemistry where the notion of asymmetric chemical evolution within chiral environments permeates a vast literature covering a wide range of topics \cite{avalos1998absolute,hananel2019enantiomeric,slkeczkowski2020}. Among many possible, we give here three illustrative examples. $(a)$ H. Kagan et al. have reported that an asymmetric synthesis can be triggered when irradiating the reactants with circularly polarized light, yielding enantiomeric excess in the product formation \cite{kagan1971photochemistry}. $(b)$ Chiral liquid crystals nuclear magnetic resonance (NMR) and $(c)$ chiral chromatography both exploit the fact that in chiral solvants, solute-solvant interactions are enantioselective. In NMR, these interactions lead to differential orientations of molecular enantiomers with respect to the magnetic field. As a consequence, NMR parameters, such as chemical shift anisotropy, are enantiomerically dependent, achieving high-resolution chiral discrimination capacities \cite{sarfati2000theoretical,lesot2015enantiotopic}. In chiral chromatography too, the chirality of the stationary phase (the chiral selector) is crucial for forming, through non covalent interactions between the enantiomers and the chiral selector, diastereoisomer complexes that have different free energies depending on the enantiomer. Differences in free energies lead to driving forces for retention in the column that become enantiodependent \cite{fornstedt1997thermodynamic,maier2001separation}. 

In these examples, the precise role played by the chiral coupling in the thermodynamics is not always easy to uncover explicitly. This has driven us to propose an optomechanical Brownian model that involves chiral optical forces and which aim is to formulate an enantiospecific thermodynamics explicitly, as schematized in Fig. \ref{fig1a}. We set our model in the framework of the thermal activation of a barrier crossing proven by H.A. Kramers to yield efficient diffusion models of chemical reactions \cite{Kramers1940} . As such, the so-called Kramers problem \cite{Melnikov1991} is immediately connected with the field of chiral chemistry where a great variety of chiral molecular systems do exhibit thermally activated bistability \cite{Shao1997,peters2017reaction}. The bistable potential is indeed, and even before Kramers, central to the first historical explanation of the stability of chiral molecules in the so-called Hund's paradox \cite{hund1927deutung}. Room temperature fast interconversions of enantiomers through conformational barriers are usually modeled as dynamical processes that lead, in the majority of cases, to racemic solutions. In this context, bistability provides a conceptual framework for describing such interconversions but also for investigating the possible modifications of the interconversion rates in order to favor one enantiomer with respect to the other, a process known as deracemization. 

In chemistry, finding such possibilities is important in that they can allow to form enantiomerically pure systems from racemic mixtures, a result of paramount importance in the pharmaceutical industry. But deracemization is usually a process that is entropically penalized with respect to racemization and that therefore demands stereoselective interactions able to bias the bistable dynamics \cite{amabilino2011spontaneous,palmans2017deracemisations}. There, the influence of external chiral optical fields in deracemization was early anticipated, notably by Pasteur, Le Bel and Van't Hoff in the 19th century, much later verified experimentally \cite{inoue1992asymmetric,feringa1999absolute}, and currently at the heart of the development of chiroptical molecular machines \cite{Feringa1996}. However, the thermodynamics involved in the variety of deracemization processes (by crystallization, chemistry, light, etc) is not always easy to resolve in an explicit way. The key result of our work is to evaluate the influence of the chiral coupling on the thermodynamics of the barrier crossing, providing an optomechanical analog of chiral discrimination and deracemization processes.

Similar issues torment the search for the origin of homochirality where, too, bistability plays a central role \cite{bonner1996quest,bonner1991origin,bada1995origins,siegel1998homochiral}. In this search, the concept of spontaneous mirror symmetry breaking is often invoked. Starting from enantiomorph local minima at distinct reaction coordinates, separated by a high activation barrier but degenerated in terms of (Gibbs) free energies, the scenario is to explain how a stochastic selection of one enantiomer trapped in its minimum, at the initial condition, then followed by an amplification or autocatalytic process, can lead to the full predominance of only one chirality, being \textit{dextro} for sugars or \textit{levo} for amino acids \cite{kondepudi1983chiral,blackmond2010}. Alternatively, one can move away from such stochastic scenarii by rather resorting, more faithfully to Pasteur's views of ``asymmetric forces'' \cite{Pasteur1860}, to underlying, systematic coupling mechanisms that can lift the enantiomeric free energy degeneracy and, as a consequence, lead to the asymmetry observed in biochemistry \cite{kondepudi2001chiral}. Such coupling mechanisms can rely on fundamental asymmetries, such as the violation of parity \cite{quack2008high,darquie2010progress}. They can also, and somehow more evidently, rely on the role of the environment (be it electromagnetic, chemical, geological, astronomical, so forth) which chirality determines the direction of the splitting in free energies \cite{cronin2005chirality}. 

It is the aim of the chiral optomechanical model presented in this Article to describe in details how a chiral environment can yield an enantiospecific thermodynamics within (Helmholtz) free energy landscapes.

\section{Summary of our framework}

From a dynamical viewpoint, the coupling between a chiral dipole and a chiral electromagnetic field (for instance, a circularly polarized light field) leads to specific types of optical forces, chiral in nature, that are exerted on the chiral dipole. Their general structure, in particular their reactive (conservative) and dissipative (non-conservative) nature, central to this work, is reminded in Section \ref{remind} of the manuscript. We explain the coupling mechanism through which new chiral optical forces emerge, as Pasteurian forces, from the ``immersion'' of the chiral dipole within a chiral electromagnetic environment. For the reasons exposed in the Introduction, we set this optical chiral coupling in a specific dual-beam optical configuration designed in such a way as to form a bistable optical trapping potential capable of implementing a Brownian thermal activation process. The dual-beam configuration and the associated double-well potential energy with its barrier height profile, are detailed in Section \ref{sec:bistable}. 

Our concept of tailored chiral optical environments is presented in Section \ref{sec:chienv} where we show that, keeping the electromagnetic energy density fixed, a fine polarization control enables to induce chiral densities and fluxes within the dual-beam interfering pattern in ratios fully controllable by the polarization parameters. We show in this Section how different types of chiral optical environments, of reactive and/or dissipative nature, can be generated depending on whether chiral densities or fluxes are selected. The nature of the environment then lead to the induction of specific chiral optical forces, that are conservative or non-conservative in reactive or dissipative chiral optical environments, respectively.

Section \ref{sec:fokker} builds the stochastic model of a Brownian chiral nanoparticle optically trapped within a bistable potential energy in order to evaluate the thermodynamics consequences of the chiral coupling in both reactive and dissipative cases. This Section presents the one-dimensional Fokker-Planck description we use in order to calculate escape rates from both sides of the potential barrier within the Kramers framework \cite{Kramers1940,hanggi1990reaction}. 

In this thermodynamics framework, Section \ref{sec:reac} calculates the modification of the escape rates induced by immersing the chiral nanoparticle inside a reactive chiral environment. Induced chiral forces, being conservative, contribute to a global change of the Helmholtz free energy landscape, yielding new potential energy surfaces that depend both on the dipole enantiomer and on the optical field enantiomorph. In contrast, when the chiral coupling is set to be dissipative in Section \ref{sec:diss}, the chiral optical forces exerted on the chiral dipole are non-conservative and the mechanical energy transferred by the chiral optical field to the dipole that is dissipated as heat into the thermal bath. In this case, the probability density function associated with the barrier crossing diffusion is modified, not by a change in the Helmholtz free energy as in the reactive case, but rather by creation of entropy. This is an important consequence of the dissipative coupling, giving in the context of chirality a clear illustration of the difference already stressed by C. Jarzynski in stochastic thermodynamics \cite{jarzynski2007comparison} between the ``inclusive work'' of reactive coupling where a conservative chiral force enters the definition of the free energy- and the ``exclusive work'' of the dissipative coupling where the symmetry breaking induced by a non-conservative chiral force is accompanied by the creation of heat/entropy, i.e. proceeding from a nonequilibium steady state configuration.

The relevance of the Kramers framework for extracting the thermodynamics significance of our chiral optomechanical model is confirmed in Section \ref{sec:simuEns} by three-dimensional stochastic simulations of an overdamped Langevin equation. Physically, Fokker-Planck and Langevin equations are equivalent but they constitute different tools. The Langevin approach gives indeed access to the individual Brownian trajectory for the diffusion of a chiral nanosphere inside the bistable optical trap in the presence of either reactive or/and dissipative chiral optical forces. Simulating the diffusive motions of a large number of chiral nanospheres leads to build statistical ensemble that greatly help visualizing the mechanical action of the chiral optical environment when calculating the probability density functions associated with the ensemble of trajectories. Consolidating our model approach based on escape rates, the simulations highlight the chiral thermodynamic deracemization scheme enabled by the dissipative chiral coupling as a potential chiral resolution strategy. 

The results are complemented in Section \ref{sec:simuSin} with statistics performed on the barrier crossing events at the level of single trajectories. These statistics lead to precise determinations of the average residency times in each of the bistable local wells. They actually point to the possibility to measure the chiral optical forces biasing the thermodynamics of the barrier crossing by only recording residency times. Exchanging thereby force measurements into time measurements paves the way to shortcut force calibration procedures and as such, to a strategy interesting to develop in the context of weak chiral force measurements. In particular, obtaining the average residency times in both the reactive and dissipative chiral coupling schemes yields an \textit{absolute} measurement of both real and imaginary parts of the chiral polarizability of a single nanoparticle. This capacity is rooted in the fact that chiral coupling transforms chiral internal degrees of freedom of the trapped Brownian object into true thermodynamic control parameters. This feature constitutes a central outcome of our work.

\section{A reminder on achiral and chiral optical forces}  \label{remind}

We remind here the general expression of the time-averaged optical force ${\bf F}$ exerted by a harmonic electromagnetic complex field $(\myvec{E}(\myvec{r})e^{-i\omega t},\myvec{H}(\myvec{r})e^{-i\omega t})$ on a chiral dipole characterized by electric $\myvec{p}$ and magnetic $\myvec{m}$ dipolar moments coupled to the incident electric and magnetic fields through complex electric $\alpha$, magnetic $\beta$ and mixed electric-magnetic $\chi$ polarizabilities as \cite{canaguier2013mechanical,BarronBook}:
\begin{eqnarray}
\left( \begin{array}{c} \myvec{p} \\ \myvec{m} \end{array} \right) =  \left( \begin{array}{cc} \alpha \varepsilon_f &    i\chi \sqrt{\varepsilon_f \mu_f}   \\ -i\chi   \sqrt{\varepsilon_f / \mu_f} & \beta \end{array} \right)   \left( \begin{array}{c}\myvec{E} \\ \myvec{H} \end{array} \right),
\end{eqnarray}
where $\varepsilon_f, \mu_f$ are the permittivity and permeability of the fluid enclosed in the optical trapping cell (deionized water). 

As now well-known, the time-averaged optical force ${\bf F}$ splits into a (standard) achiral and (new) chiral contributions. We defer to Appendix \ref{Appendix:Force_model} the full expression for ${\bf F}$ and neglect here and below all magnetic force contributions. This allows us writing simply 
\begin{eqnarray}
{\bf F}(\myvec{r}) &=& {\rm Re}\left[ \alpha {\bf f}_0 (\myvec{r})+ \frac{\chi}{\omega\sqrt{\varepsilon_f \mu_f}} {\bf h}_0(\myvec{r})\right] \notag\\ 
                   &=& {\bf F}_\alpha (\myvec{r})+ {\bf F}_\chi (\myvec{r}) ,   \label{eq:chifor}
\end{eqnarray}
where ${\bf F}_\alpha$ is the standard achiral optical force contribution that only involves $\alpha$ and ${\bf F}_\chi$ is the new chiral optical force contribution that depends on the mixed electric-magnetic $\chi$ polarizability. Both achiral and chiral force contributions can be separated into reactive and dissipative components \cite{stenholm1986semiclassical,canaguier2013mechanical} engaging respectively the real and imaginary parts $(i)$ of the $(\alpha,\chi)$ polarizabilities  and $(ii)$ of the vector fields $({\bf f}_0,{\bf h}_0)$ that take simple forms with 
\begin{eqnarray}
{\rm Re}\left[ {\bf f}_0(\myvec{r})\right] &=& \nabla W_E(\myvec{r}) \\
{\rm Im}\left[ {\bf f}_0(\myvec{r})\right] &=& -\omega\varepsilon_f\mu_f{\bf \Pi}_0(\myvec{r})  \\
{\rm Re}\left[ {\bf h}_0(\myvec{r})\right] &=& \nabla K(\myvec{r})  \\
{\rm Im}\left[ {\bf h}_0(\myvec{r})\right] &=& -2\omega \varepsilon_f \mu_f (\myvec{\Phi}(\myvec{r})-\nabla\times\myvec{\Pi}(\myvec{r})/2).
\end{eqnarray}

The achiral contribution therefore is determined by $W_E(\myvec{r}) = \varepsilon_f \myvec{E}(\myvec{r})\cdot\myvec{E}^*(\myvec{r})/4$ the time-averaged (electric) energy density and ${\bf \Pi}_0(\myvec{r})$ the orbital part of the full Poynting vector ${\bf \Pi}={\rm Re}\left[ \myvec{E}(\myvec{r})\times\myvec{H}^*(\myvec{r})\right]/2$, showing how the reactive achiral force component can be interpreted as a gradient force and the dissipative one as a radiation pressure, as already discussed in \cite{ruffnerPRL2013,canaguier2013force}. For the chiral contribution, the chiral density $K(\myvec{r})=\omega \varepsilon_f \mu_f {\rm Im}\left[ \myvec{E}(\myvec{r})\cdot\myvec{H}^*(\myvec{r})\right]/2$ and the chiral flux $\myvec{\Phi}(\myvec{r})=-\omega {\rm Im}\left[ \varepsilon_f \myvec{E}(\myvec{r})\times\myvec{E}^*(\myvec{r})+\mu_f \myvec{H}(\myvec{r})\times\myvec{H}^*(\myvec{r})\right]/4$ measure the chirality of the electromagnetic field. Here, we stress that $K(\myvec{r})$ and $\myvec{\Phi}(\myvec{r})$ are time-even, parity-odd quantities, therefore \textit{truly} chiral according to Barron's definition \cite{BarronLincei}.

These remarkable expressions reveal new types of optical forces, chiral in nature, that are induced when a chiral system is immersed within an electromagnetic field that contains either non-zero electromagnetic chiral density or chiral flux. These chiral optical forces intertwine the chirality of the matter with the chirality of the electromagnetic field and are enantioselective, explaining why they generated a strong interest since their predictions. Dipolar, they also do not depend on any specific energy-level structure of the chiral system involved. However, essentially because $\chi\ll\alpha$, chiral optical forces remain small compared to achiral optical forces. This issue has driven many proposals for exploiting the potential of these chiral optical forces in chiral discriminatory schemes despite the fact that they correspond to relatively weak signals \cite{canaguier2013mechanical,cameron2014diffraction,ding2014realization,tkachenko2014optofluidic,canaguier2015chiral,Hayat13190,rukhlenko2016completely,dionne2017nano,kravets2019optical}. 

The chiral light-matter coupling leads to simple relations. From the light part, chiral electromagnetic fields form a pair of enantiomorph optical environments when reversing the signs of $K(\myvec{r}),\myvec{\Phi}(\myvec{r})$ without changing the energy density. From the matter part, chiral dipoles form a pair of enantiomers with opposite signs for the real and imaginary parts of $\chi$. We decide here to call a ``right-handed'' dipole one with ${\rm Re}\left[\chi\right] >0, {\rm Im}\left[\chi\right] <0$ and a ``left-handed'' dipole with ${\rm Re}\left[\chi\right] <0, {\rm Im}\left[\chi\right] >0$. In the model presented below, we fix a ratio $\chi /\alpha =5\%$ calculated from the Clausius-Mossotti polarizabilities $\alpha$ and $\chi$ in the quasistatic limit --see Appendix \ref{Appendix:polarizabilities} for details. 

\section{Bistable potential energy in an optical trap}  \label{sec:bistable}

\begin{figure}[htb!]
  \centering{
    \includegraphics[width=0.9\linewidth]{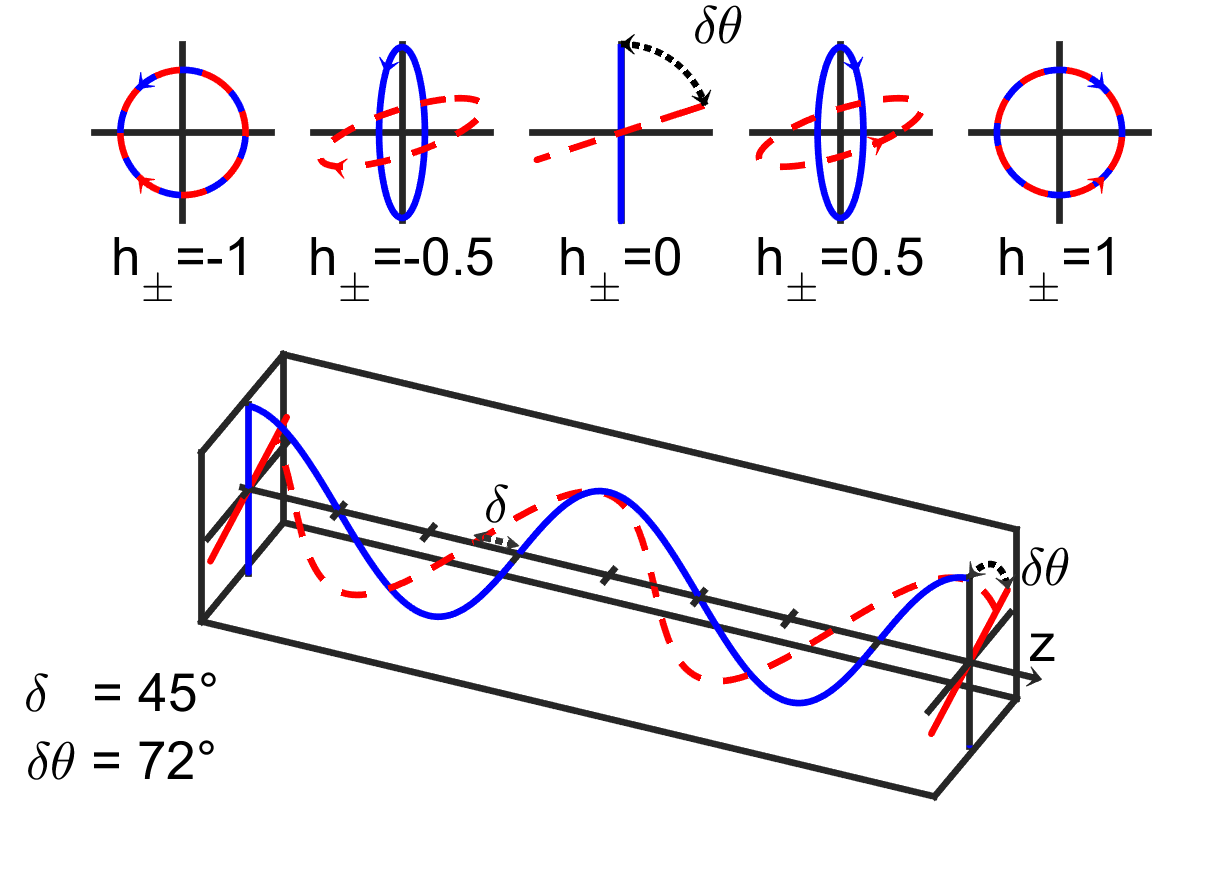}
    }
  \caption{The polarization vectors $\myvec{e}_\pm$ for each of the two counter-propagating Gaussian beams are represented for plane wave electric fields. In blue, a beam with $\myvec{e}_+$ and $k>0$ and in red, with $\myvec{e}_-$ and $k<0$. The schematics illustrates the effects of the phase $\delta$ and the polarization main axis angle $\delta\theta$ on beams linearly polarized with $h_\pm=0$. The insets show the polarization ellipses for different values of $h_\pm$. The effect of the $\delta\theta$ parameter is seen when the beams are not circularly polarized.  }
  \label{Fig:PolaParam}
\end{figure}

The expressions of the optical forces being reminded, we now explain how the achiral force contribution can induce a bistable dynamics within the optical trap. To do so, i.e. to form a double-well trapping potential, we use a trapping configuration involving two counter-propagating Gaussian beams, focused on a common waist, already implemented in the context of optical force spectroscopy \cite{ashkin1970,Smith795,vanderHorst:08}. In the paraxial approximation \cite{varga1998gaussian} and using harmonic time dependent complex fields, the Gaussian beams, propagating either with a $+kz$ or $-kz$ phase along the $z$-optical axis ($k=\sqrt{\varepsilon_f}\omega /c$), can be evaluated at any position $\myvec{r}=q\hat{\boldsymbol{\rho}}+z\hat{\bf z}$ in the cylindrical coordinate system as:
\begin{eqnarray}
\myvec{E}_{\pm}(\myvec{r})&=&\mathcal{E}_{\pm}(\myvec{r})e^{{\pm}ikz} \ \myvec{e}_{\pm} ,  \\
\myvec{H}_{\pm}(\myvec{r})&=&\frac{1}{Z_{f}}\mathcal{E}_{\pm}(\myvec{r})e^{{\pm}ikz}\left(\pm\hat{\bf z}\times\myvec{e}_{\pm}\right),
\end{eqnarray}
where $\myvec{e}_{\pm}$ are the (unit) polarization vectors associated with each field in each direction of propagation and $Z_{f}=\sqrt{\mu_f/\varepsilon_f}$ the optical impedance of the fluid. With beam waists $w_0$ and Rayleigh ranges $z_R$ identical for both beams, we have
\begin{eqnarray}
\mathcal{E}_{\pm}(\myvec{r})&=&\mathcal{E}_0\frac{w_0}{w(z)}e^{{\pm}i\phi(q,z)}e^{-\frac{q^2}{w^2(z)}}
\end{eqnarray}
where we note $\phi(\myvec{r})=k q^2/2R(z)-\xi(z)$ the Gaussian phase that accounts for the finite radius of curvature $R(z)=z[1+(z_R/z)^2 ] $ of the beam and the Gouy phase $\xi(z)=\arctan [z/z_R]$, and $w(z)=w_0[1+(z/z_R)^2 ] ^{1/2}$ the beam radius measured along the optical axis from both sides of the waist.
We define the polarization vectors by 
\begin{eqnarray}
\myvec{e}_{+} &=& (\sqrt{1-h_+}\myvec{e}_{l} + \sqrt{1+h_+}\myvec{e}_{r})/\sqrt{2}  \nonumber  \\
\myvec{e}_{-} &=& (\sqrt{1-h_-}e^{i(\delta-\delta\theta)}\myvec{e}_{l} + \sqrt{1+h_-}e^{i(\delta+\delta\theta)}\myvec{e}_{r})/\sqrt{2}  \nonumber
\end{eqnarray} 
in the basis of left and right circular polarization vectors $\myvec{e}_{l}$ and $\myvec{e}_{r}$, with $h_+$ and $h_-$ corresponding to the helicity of both beams ranging from $1$ for a right-handed circular polarization to $-1$ for a left-handed circular polarization. The phase delay between both beams is $\delta$ and $\delta\theta$ is the angle between the semi major axis of the polarization of both beams, as described in Fig. \ref{Fig:PolaParam}. The field superpositions $\Esw (\myvec{r})= \myvec{E}_+ (\myvec{r})+ \myvec{E}_- (\myvec{r})$ and $\Hsw (\myvec{r})$ form a standing-wave. A crucial consequence for the forces is the zero Poynting vector inside the standing-wave because $\Esw(\myvec{r})\times\Hsw^*(\myvec{r})$ is purely imaginary. 

Before inducing any chiral coupling, let us look at the dynamical landscape within the optical trap when solely involving the achiral reactive force field ${\bf F}_\alpha (\myvec{r})={\rm Re}\left[\alpha\right] \nabla W_E(\myvec{r})$. This force is conservative and the corresponding potential energy inside the optical trap $U_{\rm opt}(\myvec{r})=-{\rm Re}\left[\alpha\right] W_E(\myvec{r})$ is determined by the time averaged electric energy density $W_E(\myvec{r}) = \varepsilon_f \Esw(\myvec{r})\cdot\Esw^*(\myvec{r})/4$ inside the standing-wave. 

The notations
\begin{eqnarray}
&&h_1 = (\sqrt{1-h_+}\sqrt{1-h_-}-\sqrt{1+h_+}\sqrt{1+h_-})/2 \notag \\
&&h_2 = (\sqrt{1-h_+}\sqrt{1-h_-}+\sqrt{1+h_+}\sqrt{1+h_-})/2 \notag \\
&&\varphi(\myvec{r}) = \delta + 2 k \left(z + \frac{q^2}{2 R(z)}\right) - 2 \xi(z)\notag
\end{eqnarray} 
allow us to express in a simple way the separation of the energy density $W_E(\myvec{r})$ between a trapping energy density $W_{trap}(\myvec{r})$ independent of polarization, and an interference energy density $W_{inter}(\myvec{r})$ according to:  
\begin{eqnarray}
W_E(\myvec{r}) &=& W_{trap}(\myvec{r}) + W_{inter}(\myvec{r}) \\
W_{trap}(\myvec{r}) &=& \frac{\mathcal{E}_0^2 w_0^2 \varepsilon_f}{2 w^2(z)} e^{-\frac{2 q^2}{w^2(z)}}\\
W_{inter}(\myvec{r}) &=& W_{trap}\left(\myvec{r}\right) \bigg(h_2 \cos\delta\theta \cos\varphi(\myvec{r}) + \notag \\
& & h_1 \sin\delta\theta \sin\varphi(\myvec{r})\bigg).
\end{eqnarray}

\begin{figure}[htb!]
  \centering{
    \includegraphics[width=0.8\linewidth]{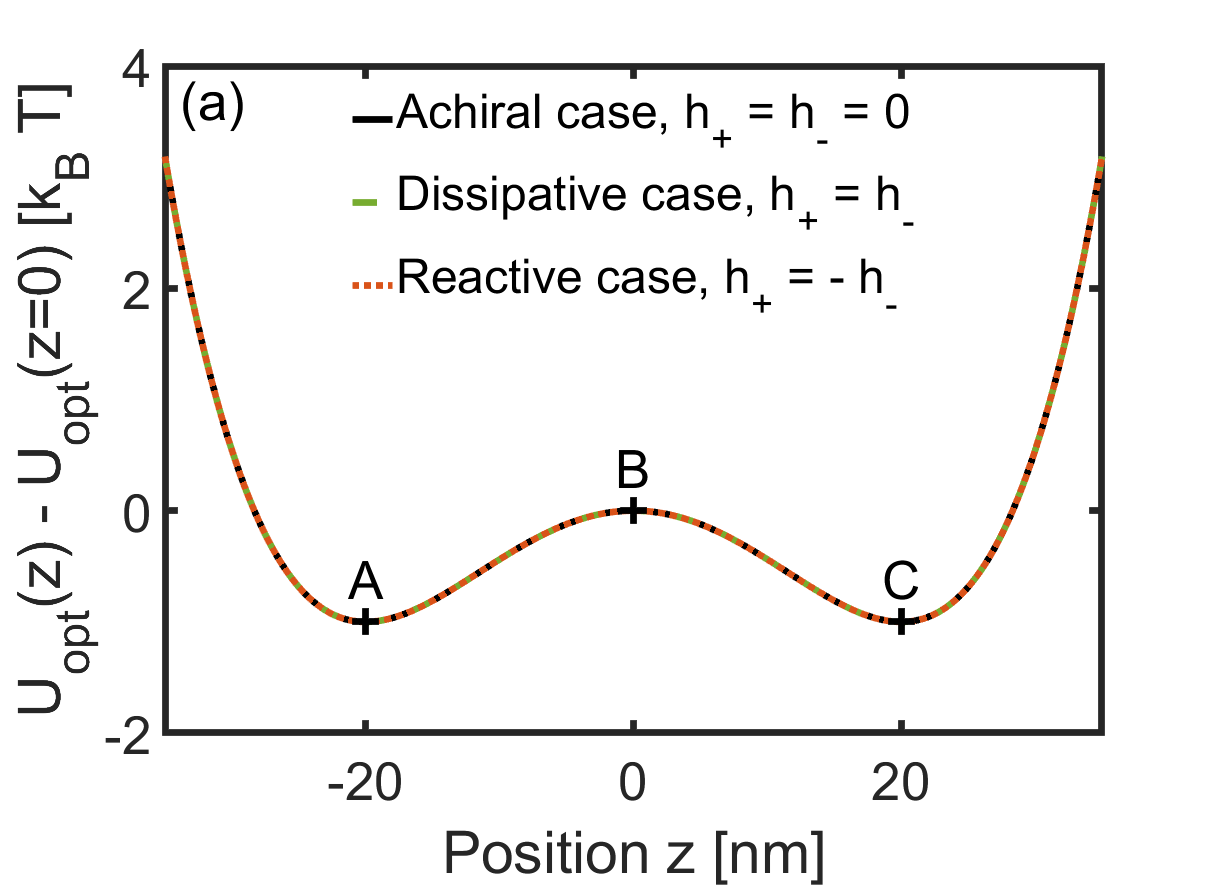}
    \includegraphics[width=0.8\linewidth]{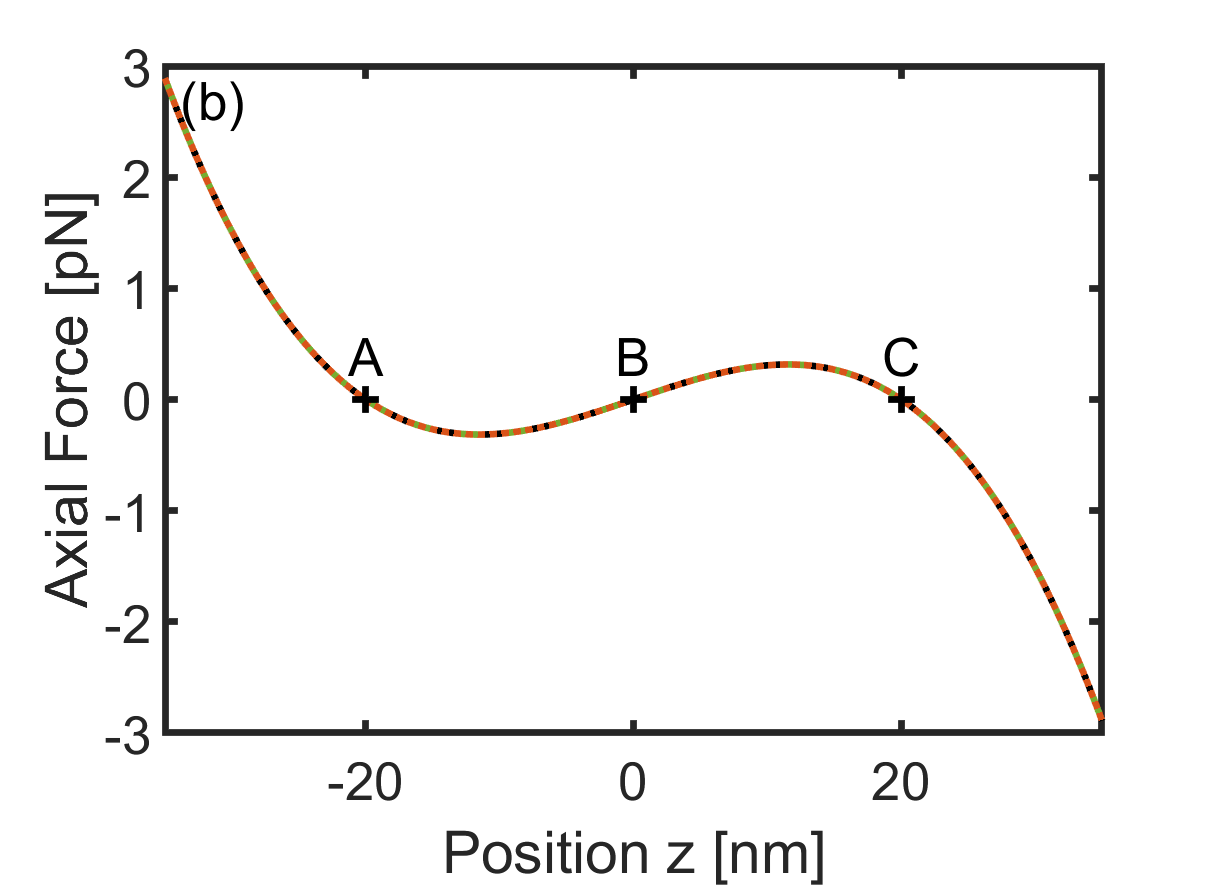}
    }
  \caption{(a) Bistable optical potential energy $U_{\rm opt}(\myvec{r})=-{\rm Re}\left[\alpha\right] W_E(\myvec{r})$ displaying the two local minima at $z_A$ and $z_C$ separated by the barrier at $z_B$. The electric dipolar polarizability $\alpha$ associated with an achiral ($\chi=0$) gold (Au) nanosphere of radius $R=\SI{20}{\nano\meter}$ is calculated using Au tabulated optical data measured at an illumination wavelength of $\SI{785}{\nano\meter}$ and the Clausius-Mossotti relations of Appendix \ref{Appendix:polarizabilities} with $\kappa_{Au} \equiv 0$ in this case. (b) Corresponding optical achiral force field ${\bf F}_\alpha={\rm Re}\left[\alpha\right] \nabla W_E(\myvec{r})$ drawn in the waist region. This double-well profile in $U_{\rm opt}(\myvec{r})$ is generated with the polarization settings $h_+=h_-=0, \delta\theta = 0.9989\times \pi/2, \delta=-\pi$. Superimposed in (a) and (b) are the achiral potential energy recalculated with the polarization settings that lead to reactive and dissipative chiral environments. These settings are $h_+=0.05=-h_-, \delta\theta = 0.8990\times \pi/2, \delta=-\pi$ for the reactive case and $h_+=0.0017=h_-, \delta\theta = \pi/2, \delta=\pi/2$ for the dissipative one. As seen, these polarization settings do not modify the achiral potential energy surface inside the optical trap. The distance between the two local wells located at $z_A<0$ and $z_C>0$ from both sides of the waist barrier positioned at $z_B=0$ is noted $\Delta\ell$ in the main text. }
  \label{Fig:AchiralPotential}
\end{figure}

There is clearly a vast $(h_1,h_2,\delta,\delta\theta)$ parameter space available for the design of the potential energy landscape, as discussed in details in Appendix \ref{Appendix:Force_model}. In order to set the double-well trapping potential, we start with linear polarizations $h_+=h_-=0$ giving $h_1=0,h_2=1$. The bistability profile can then be shaped by controlling the strength of the interferences superimposed to the trapping potential. This is done by adjusting $\delta\theta$ to a value that leaves only one interference falling inside the trapping envelop strong enough to cause a force inversion around the waist --see Fig. \ref{Appendix:Fig:EHForceLandscapeZoom} in Appendix \ref{Appendix:Force_model} for a detailed description of the landscape. This being fixed, we force with $\delta$ the potential energy $U_{\rm opt}(\myvec{r})$ to be symmetric with respect to the waist position with a constructive interference at $z=0$. Finally, the barrier height is adjusted via the two beam (even) intensities. These controls lead to the bistable optical potential inside the trap displayed in Fig. \ref{Fig:AchiralPotential} (a) and (b) for the corresponding $z-$axial force field,  with the corresponding values given in the figure caption. 

\section{Bistability in chiral optical environments} \label{sec:chienv}

The explicit expressions of the electromagnetic chiral density and chiral flux associated with the dual-beam configuration described above
\begin{eqnarray}
K(\myvec{r}) &=& -(h_+ - h_-)  \cdot \omega \sqrt{\varepsilon_f \mu_f} W_{trap}(\myvec{r})  \label{eq:K}   \\
\myvec{\Phi}(\myvec{r}) &=&  -(h_+ + h_-)\cdot \omega  W_{trap}(\myvec{r})  \hat{\myvec{z}} \label{eq:Phi_EH}
\end{eqnarray}
immediately reveal that setting linear $h_+=h_-=0$ polarizations for both beams deprive the interference pattern from any chirality. But elliptically polarized beam endow the optical environment with chirality. This leads to the dynamical consequences that we now discuss. 

The first key feature of our model is the possibility to choose $h_+,h_-$ values that select $K(\myvec{r})$ or $\myvec{\Phi}(\myvec{r})$ (or both) while preserving exactly the bistable structure of the achiral potential energy defined in Sec. \ref{sec:bistable} above. This is clearly seen in Fig. \ref{Fig:AchiralPotential} (a) and (b). With such polarization choices therefore, the double-well landscape of the trap becomes optically chiral. According to Eq. (\ref{eq:chifor}), as soon as a chiral dipole is immersed in this chiral optical environment, the chiral coupling will induce chiral forces that add to the bistable dynamic which is, for its part, driven by the achiral force fields only. 

\begin{figure}[htb!]
  \centering{
    \includegraphics[width=0.9\linewidth]{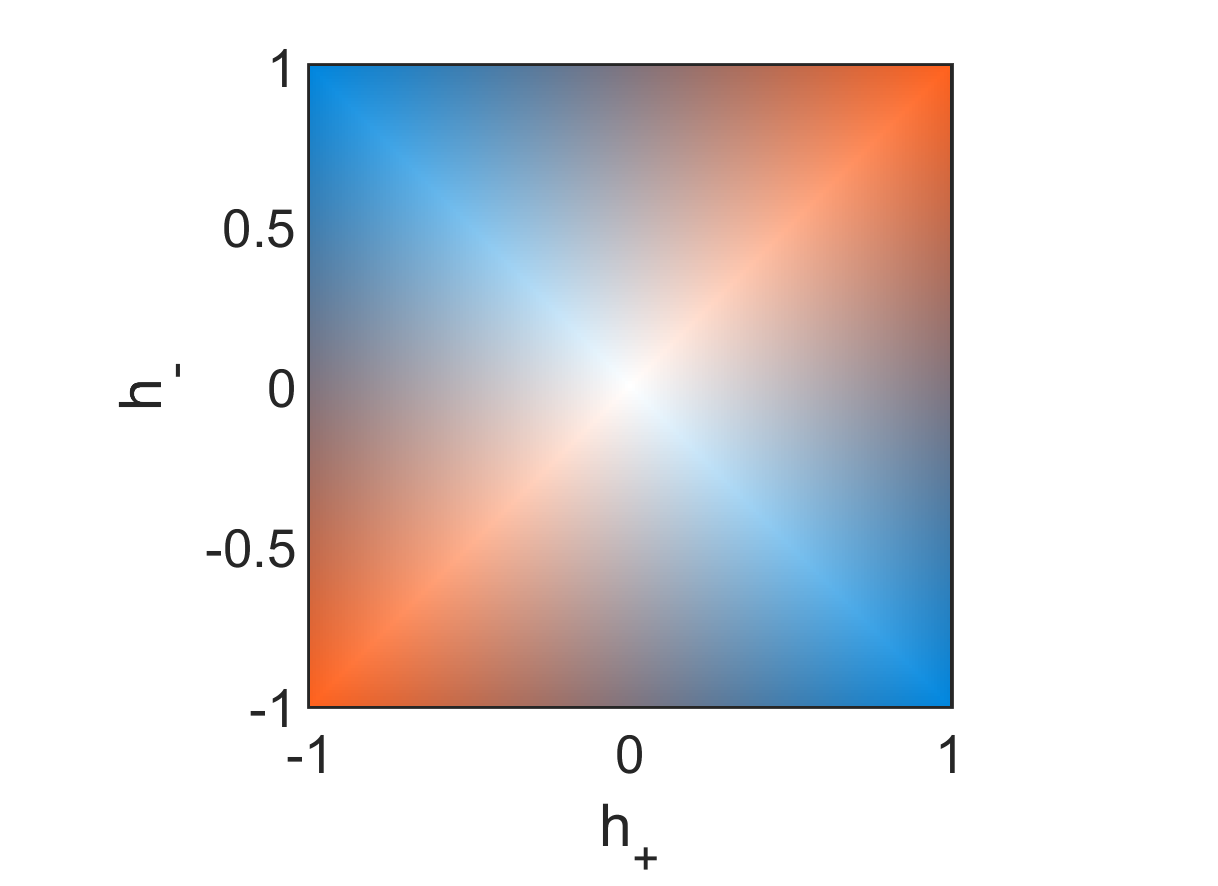}
    }
  \caption{Surface plot of the evolution of the reactive vs. dissipative nature of the chiral optical environment in the helicity plane $(h_+,h_-)$ of the two counter-propagating beams. In agreement with Eqs. \ref{eq:K} and \ref{eq:Phi_EH}, $c = \abs{h_+ - h_-}/2\cdot (c_R-c_A) + \abs{h_+ + h_-}/2\cdot (c_D-c_A) + c_A$ where $c$, $c_R$, $c_D$ and $c_A$ are respectively the displayed, blue --for reactive--, red --for dissipative-- and white --for achiral-- colors. }
  \label{Fig:ForceSpace}
\end{figure}

The second important feature is the ability to select by polarization the reactive and/or dissipative nature of the chiral environment and thereby to induce on the chiral dipole reactive and dissipative forces
\begin{eqnarray}
&&{\bf F}_\chi^{\rm reac}(\myvec{r}) = {\rm Re}[\chi]\frac{1}{\omega\sqrt{\varepsilon_f \mu_f}}\nabla K(\myvec{r})  \\
&&{\bf F}_\chi^{\rm diss}(\myvec{r}) = {\rm Im}[\chi]2\sqrt{\varepsilon_f \mu_f}\myvec{\Phi}(\myvec{r})
\end{eqnarray}
that are, each, associated with one unique chiral quantity. The evolution of the reactive vs. dissipative nature of the chiral environment in the $(h_+,h_-)$ helicity space is displayed in Fig. \ref{Fig:ForceSpace} where it is clear that the two distinct reactive $[K(\myvec{r})\neq 0,\myvec{\Phi}(\myvec{r})=\myvec{0}]$ vs. dissipative $[K(\myvec{r})= 0,\myvec{\Phi}(\myvec{r})\neq\myvec{0}]$ chiral optical environments can be selected using $h_+=-h_-$ vs. $h_+=h_-$. We stress that this selection is performed on a sole polarization control, without changing the intensity of the two beams. This polarization-based tailoring of the chiral optical environment yields the important thermodynamics consequences at the heart of our work. 

\begin{figure}[htb!]
  \centering{
    \includegraphics[width=0.7\linewidth]{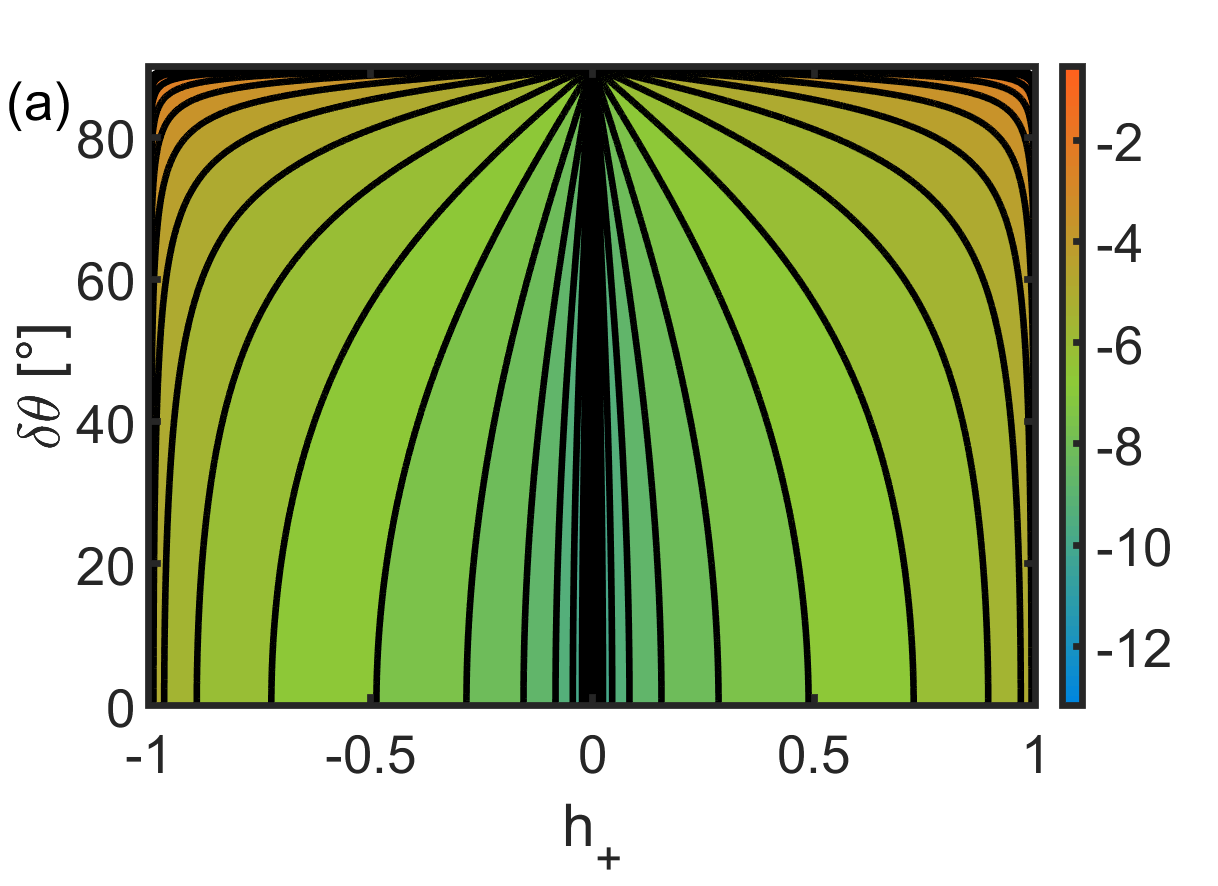}
    \includegraphics[width=0.7\linewidth]{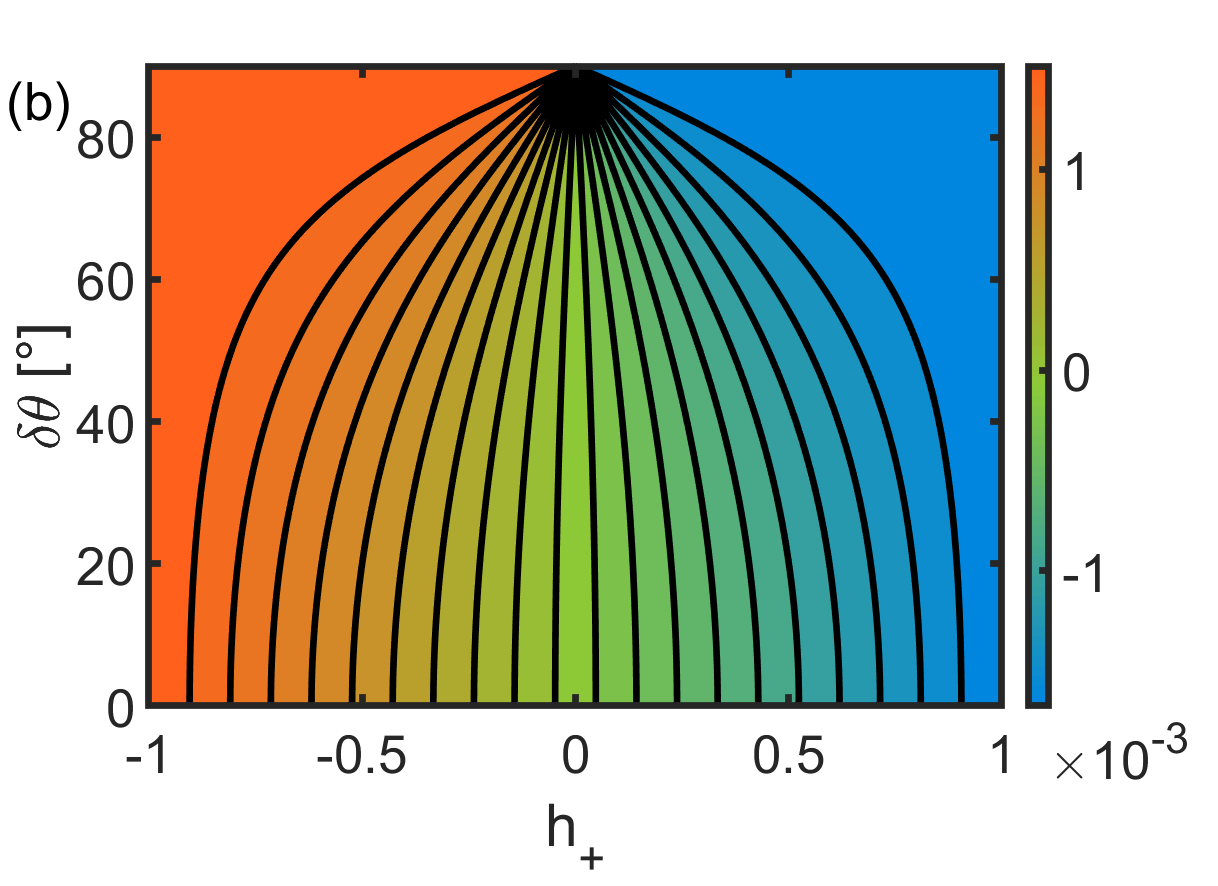}
    }
  \caption{(a) Surface plot in the $(h_+,\delta\theta)$ plane of $\log\left[{\rm Max}({\bf F}_\chi^{\rm reac}) / {\rm Max}({\bf F}_{inter})\right]$ evaluated on the optical axis in a purely dissipative chiral force configuration. (b) Surface plot in the $(h_+,\delta\theta)$ plane of ${\rm Max}({\bf F}_\chi^{\rm diss}) / {\rm Max}({\bf F}_{inter})$ evaluated on the optical axis in a purely dissipative chiral force configuration. In both reactive and dissipative couplings, $\delta\theta$ and $h_+$ are chosen in order to maximize the ratio of chiral vs. interferential forces while keeping the interferential forces sufficiently weak as to ensure the double-well structure described in Fig. \ref{Fig:AchiralPotential}. For these calculations, the intensities in the two beams, the wavelength, the nanosphere dipolar polarizability are the same as in Fig. \ref{Fig:AchiralPotential}. We set the chiral polarizability $\chi/\alpha=5\%$ as discussed in Appendix \ref{Appendix:polarizabilities}.}
  \label{Fig:ForceRatios}
\end{figure}

Importantly, the appropriate choices of polarizations that induce chirality without perturbing the achiral bistable potential energy set above in Section \ref{sec:bistable} must balance two potentially competing constraints. First, they must comply with the necessity to keep the achiral potential energy unmodified that, as discussed above, demands to decrease the amplitude of the interferences sufficiently so that the optical potential takes a double-well profile at the minimum of $W_{trap}(\myvec{r})$. Then, because chiral optical forces are weak signals, polarization settings have to allow for an optimal ratio between chiral and interferential axial forces, the latter corresponding to ${\bf F}_{inter}={\rm Re}\left[\alpha\right] \nabla W_{inter}(\myvec{r})$. The ratio associated with chiral reactive and dissipative forces are plotted in Fig. \ref{Fig:ForceRatios} (a) and (b), respectively, in the $(h_+,\delta\theta)$ parameter plane, considering that $\delta$ is tuned to shape an achiral potential energy surface symmetrical with respect to the plane $z=0$. For the reactive coupling that involves $K(\myvec{r})$, the optimal choice would be to set $h_+=-h_-=\pm 1$ with $\delta\theta=\pi /2$. However in this case, no interference is expected, losing therefore the double-well structure. This demands to slightly move away from $\delta\theta=\pi /2$ while reducing the helicity of the two beams. In contrast, the dissipative coupling involves $\myvec{\Phi}(\myvec{r})$ where $h_+=h_-$ maximizes interferential forces associated with very deep wells in the potential energy. Our choice here is rather to set $\delta\theta=\pi /2$ with a reduced helicity in both beams. These constraints lead to the polarization choices detailed in the caption of Fig. \ref{Fig:AchiralPotential} that yield force ratios strong enough for our purposes while preserving the double-well profile of $W_{E}(\myvec{r})$. 

\section{Fokker-Planck model for thermal activation of a barrier crossing: bistable equilibrium}
\label{sec:fokker}

We now include temperature $T$ and describe the evolution of the dipole inside the bistable potential, at first without chiral contributions $\left[K(\myvec{r}) =0,\myvec{\Phi}(\myvec{r})={\bf 0}\right]$. In this case, the evolution is only driven by decoupled achiral reactive axial ${\bf F}_\alpha (\myvec{r})\cdot \hat{\myvec{z}}$ and radial ${\bf F}_\alpha (\myvec{r})\cdot \hat{\boldsymbol{\rho}}$ forces inside the optical trap. This situation corresponds to a Kramers problem with the possibility given to the dipole to escape local trapping sites by thermal activation and diffusion over the separating barrier of the double-well potential energy landscape drawn in Fig. \ref{Fig:AchiralPotential}. We model this metastable dynamics with an overdamped Fokker-Planck equation \cite{hanggi1990reaction}
\begin{eqnarray}
\partial_t p(\myvec{r},t)=-\myvec{\nabla} \cdot \myvec{j}(\myvec{r},t) ,
\end{eqnarray}
connecting the probability density $p(\myvec{r},t)$ to find the dipole at $\myvec{r}$ at time $t$ to the probability current
\begin{eqnarray}
\myvec{j}(\myvec{r},t)=-\frac{1}{\gamma}\myvec{\nabla} U_{\rm opt}(\myvec{r})p(\myvec{r},t)-D\myvec{\nabla} p(\myvec{r},t),
\end{eqnarray} 
with $\gamma$ the Stokes drag, $k_{\rm B}$ the Boltzmann constant, and $D = k_{\rm B} T / \gamma $ the free Brownian diffusion coefficient. For an optical trap immobilizing an Au nanosphere of radius $R=20$ nm in pure water at room temperature ($\gamma=2\pi\eta R$ with a viscosity $\eta=0.88\times 10^{-3}$ Kg/m/s), the overdamped regime is well reached with a momentum relaxation time given by the nanoparticle over friction ratio of $m/\gamma \sim 10^{-7}$ s.

In our model, we will only study the component $j_z(\myvec{r},t)$ of the probability current along the optical $z-$axis in the steady-state regime with $U_{\rm opt}$ time-independent. In this regime, $\partial_t p(\myvec{r},t)=0$ implies that
\begin{eqnarray}
\frac{1}{q}\partial_q (q j_\rho(\myvec{r},t)) + \partial_z j_z(\myvec{r},t)=0 .
\end{eqnarray}
We now further neglect the transverse variations of the beam with respect to the axial ones, i.e. $\partial_q (q j_\rho) / q \ll \partial_z j_z$, so that the $\myvec{z}$-component of the probability current around the waist is modeled as a constant $j_z(q)$ in the $z$ variable. This makes it easy to evaluate the crossing rates for the dipole over the barrier (positioned at $z_B=0$) in the forward $+$ and backward $-$ $z-$directions. In the forward $z-$direction, the rate to be computed corresponds to crossing events from the $z<0$  initially populated well (well A, minimum at $z_A$) towards the $z>0$ unoccupied one (well C, minimum at $z_C$). This initial population corresponds to the stationary nonequilibrium probability density $p^+(q,z)$ inside well A that creates a current $j^+_z(q)=\myvec{j}(q)\cdot\hat{\myvec{z}}$ flowing in the $+z$ direction, according to: 
\begin{eqnarray}
j_z^+(q)&=&-\frac{1}{\gamma}\partial_z U_{\rm opt}(q,z)p^+(q,z) \nonumber  \\
&&-D\partial_z p^+(q,z),
\end{eqnarray} 
%

Following a standard method \cite{hanggi1986escape}, the nonequilibrium probability density $p^+(q,z)$ can be determined between one point $z^\prime$ within well A and a distant point above the barrier for which $p^+(q,z^+>z_B)=0$, as
\begin{eqnarray}
p^+(q,z^\prime )= \gamma j^+_z(q) e^{\frac{-U_{\rm opt}(q,z^\prime)}{k_{\rm B} T}}\int\limits_{z^\prime}^{z^+}{\rm d}u \ e^{\frac{U_{\rm opt}(q,u)}{k_{\rm B} T}},\label{eq:pPlusBarrierInt}
\end{eqnarray} 
together with the corresponding population density
\begin{eqnarray}
n^+_A=\int\limits_0^{+\infty}\dd{q}{2\pi q}\int\limits_{-\infty}^{z_B} {\rm d}z^\prime \ p^+(q,z^\prime).  \label{eq:pop}
\end{eqnarray}


This population is then evaluated with a Gaussian steepest-descent approximation, expanding the optical potential around the barrier maximum at $z_B$ and the local minimum at $z_A$:
\begin{eqnarray}  
{U}_{\rm opt}(q,z\sim z_B)&\simeq& {U}_{\rm opt}(q,z_B) - {b}^2(q) (z-z_B)^2  \\
{U}_{\rm opt}(q,z\sim z_A)&\simeq& {U}_{\rm opt}(q,z_A) + {a}^2(q) (z-z_A)^2 
\end{eqnarray} 
with $2{b}^2(q)=|\partial^2{U}_{\rm opt}(q,z)/\partial z^2|_{z_B}$ and $2{a}^2(q)=|\partial^2{U}_{\rm opt}(q,z)/\partial z^2|_{z_A}$, and extending the lower and upper limits of integration to $\pm\infty$. We thus obtain
\begin{eqnarray}
{n^+_A} \simeq {\int\limits_0^{+\infty}\dd{q}2\pi q \frac{k_{\rm B} T \gamma \pi j^+_z(q)}{{a}(q){b}(q)} e^{\frac{U_{\rm opt}(q,z_B)-U_{\rm opt}(q,z_A)}{k_{\rm B} T}}}.
\end{eqnarray} 

Due to the axial symmetry of the optical landscape, it is clear that the optical potential is an even function of $q$. In the close vicinity of the optical axis therefore, one can always assume that $\pdv{U_{\rm opt}}{q}{z}\relax(q, z) \sim 0$. This assumption has two consequences: $(i)$ that ${a}(q) \sim {a}$ and ${b}(q) \sim {b}$ are independent of $q$, and $(ii)$ that $U_{\rm opt}(q, z_B)-U_{\rm opt}(q, z_A) \simeq U_{\rm opt}(0, z_B)-U_{\rm opt}(0, z_A)$ by expanding the potential energy around $z_A$ and $z_B$. Under this hypothesis therefore: 
\begin{eqnarray}
n^+_A \simeq  \frac{k_{\rm B} T \gamma \pi} {{a} {b}} e^{\frac{U_{\rm opt}(0,z_B)-U_{\rm opt}(0,z_A)}{k_{\rm B} T}}\int\limits_0^{+\infty}\dd{q}2\pi q j^+_z(q).
\end{eqnarray} 
The assumption leads to interpret $\int\limits_0^{+\infty}\dd{q}2\pi q j^+_z(q) = J^+_z$ as the total probability current in the positive direction. The escape rate $\kappa_{A\rightarrow C}$ from well A to well C then simply writes as
\begin{eqnarray}  
\kappa_{A\rightarrow C} = \frac{J^+_z}{n^+_A} \simeq   \frac{a{b}}{k_{\rm B} T \gamma\pi}e^{-\frac{U_{\rm opt}(0,z_B)-U_{\rm opt}(0,z_A)}{k_{\rm B} T}}, \label{eq:rateA}
\end{eqnarray} 
which corresponds to the well-known result obtained by Kramers with $\Delta U^{AB}_{\rm opt}=U_{\rm opt}(0,z_B)-U_{\rm opt}(0,z_A)$ the optical barrier height measured along the optical axis at $q=0$ \cite{Kramers1940,hanggi1990reaction} .

The escape rate $\kappa_{C\rightarrow A}$ from well C to well A is calculated from the probability current $j^-_z(q)=\myvec{j}^-(q)\cdot (-\myvec{z})$, flowing in the opposite direction than $j^+_z(q)$ and solution of
\begin{eqnarray}
j_z^-(q)&=&+\frac{1}{\gamma}\partial_z U_{\rm opt}(q,z)p^-(q,z) \nonumber  \\
&&+D\partial_z p^-(q,z).
\end{eqnarray} 
Following the same steps, but this time integrating over well C, one evaluates the escape rate from the well C at $z_C$ over the barrier at $z_B$ as:
\begin{eqnarray}  
\kappa_{C\rightarrow A}= \frac{J^-_z}{n^-_C}\simeq  \frac{{b}{c}}{k_{\rm B} T \gamma\pi}e^{-\frac{U_{\rm opt}(0,z_B)-U_{\rm opt}(0,z_C)}{k_{\rm B} T}}.  \label{eq;rateC}
\end{eqnarray} 

For a symmetric optical potential with $U_{\rm opt}(0,z_A)=U_{\rm opt}(0,z_C)$ and ${a}={c}$, one obviously obtains $n^+_A=n^-_C$ and therefore from the detailed balance $\kappa_{A\rightarrow C} =\kappa_{C\rightarrow A}$. We can take this equality and the absence of any other force besides the trapping force forming the optical bistable potential energy as the definition of the equilibrium state of our system. 

\section{Steady-state in the reactive chiral coupling}
\label{sec:reac}

We select here the polarizations $h_+=h_-$ in the two beams in order to induce a purely reactive chiral coupling and to study its impact on the thermodynamics of the thermal activation process inside the bistable optical trap. In this case, the reactive chiral optical force derives from the gradient of the chirality density and is thus conservative. It therefore contributes to the optical energy potential as a chiral potential 
\begin{eqnarray}
U_{\chi}(q,z)= -{\rm Re}\left[\chi\right] K(q,z) / \omega\sqrt{\varepsilon_f\mu_f}
\end{eqnarray} 
that adds to the dynamics described by the steady-state Fokker-Planck equation, according to (forward direction)
\begin{eqnarray}
\widetilde{j}^+_z(q)&=&-\frac{1}{\gamma}\partial_z U_{\rm pot}(q,z)\widetilde{p}^+(q,z) \nonumber  \\
&&-D\partial_z \widetilde{p}^+(q,z),
\end{eqnarray} 
defining $U_{\rm pot}(q,z)=U_{\rm opt}(q,z)+U_{\chi}(q,z)$. With a modified $\widetilde{p}^+$ nonequilibrium probability density, escape rates evolve accordingly with:
\begin{eqnarray}  
\widetilde{\kappa}_{A\rightarrow C}&=& \frac{\widetilde{J}^+_z}{\widetilde{n}^+_A}\nonumber  \\
&&\simeq  \kappa_{A\rightarrow C}\cdot e^{-\frac{U_{\chi}(0,z_B)-U_{\chi}(0,z_A)}{k_{\rm B} T}} , \label{rates:reac1}\\  
\widetilde{\kappa}_{C\rightarrow A} &=& \frac{\widetilde{J}^-_z}{\widetilde{n}^-_C}\nonumber  \\
&&\simeq  \kappa_{C\rightarrow A} \cdot e^{-\frac{U_{\chi}(0, z_B)-U_{\chi}(0, z_C)}{k_{\rm B} T}},\label{rates:reac2}
\end{eqnarray} 
where we have verified that the local curvatures of the optical landscape are only weakly modified by the chiral potential, in other words that $\widetilde{a}\simeq a$, $\widetilde{b}\simeq b$ and $\widetilde{c}\simeq c$. We use the same notation for $U_\chi$ used for $U_{\rm opt}$ above, and where we take advantage of the $z-$parity of the chiral density $K(q,z)$ with $\partial_z U_\chi (z)|_{z=0}=0$ and its $q-$parity giving $\pdv{U_{\rm opt}}{q}{z}\relax(q, z) \sim 0$ close to the optical axis. This parity also implies that the reactive chiral coupling does not lift the degeneracy in free energy between the two wells, maintaining the equilibrium constant to $\widetilde{\kappa}_{A\rightarrow C}/\widetilde{\kappa}_{C\rightarrow A}=1$. 

From a thermodynamics viewpoint, the rate modifications come from the work performed by the reactive chiral force between the barrier and the wells. This \textit{conservative} work provides a contribution to the potential energy in the form of a Helmholtz free energy difference $\Delta \mathcal{F}_{\chi i} = W_{\chi i}^{\rm cons}=U_{\chi}(z_B)-U_{\chi}(z_i)$, with $i=A,C$. This situation exactly corresponds to a chiral counterpart of the inclusive framework discussed by Jarzynski \cite{jarzynski2007comparison}. 

\begin{figure}[htb!]
  \centering{
    \includegraphics[width=0.8\linewidth]{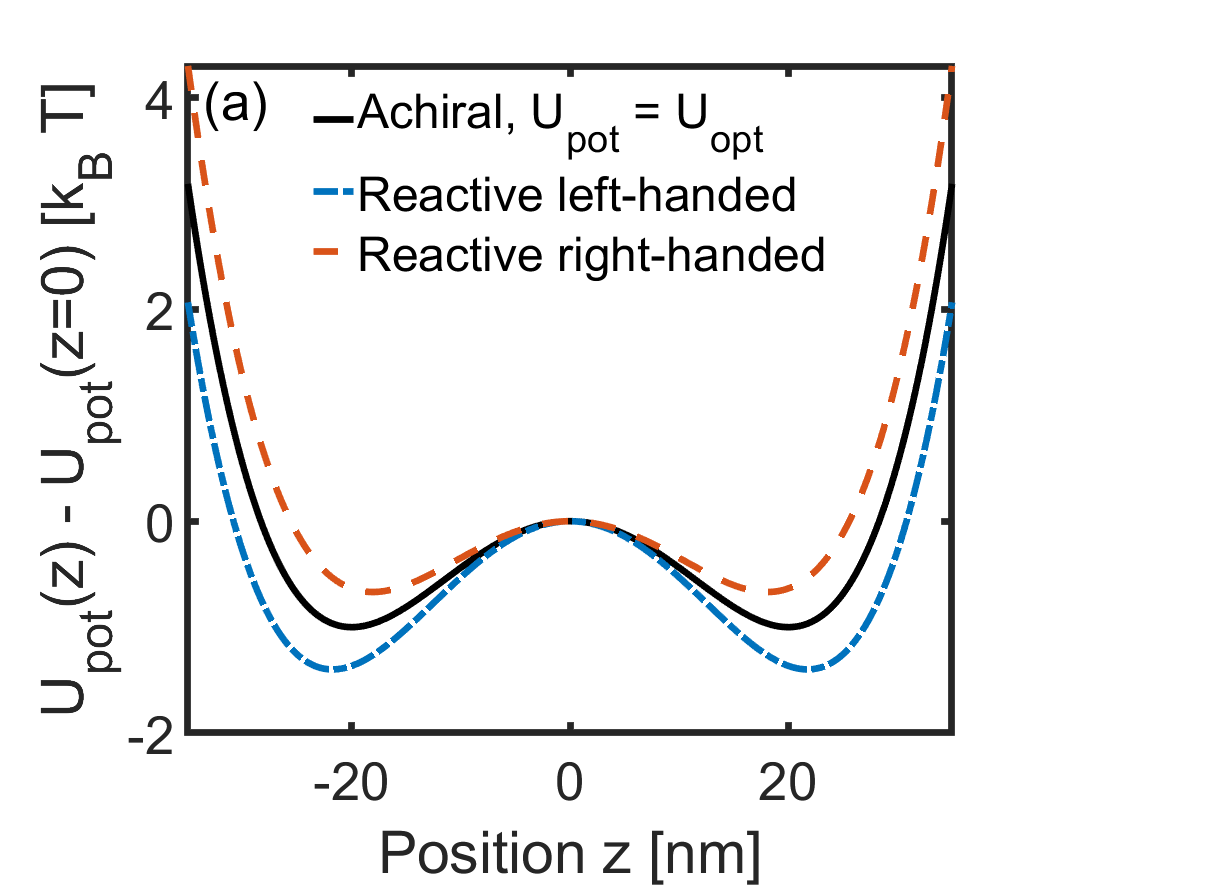}
    \includegraphics[width=0.8\linewidth]{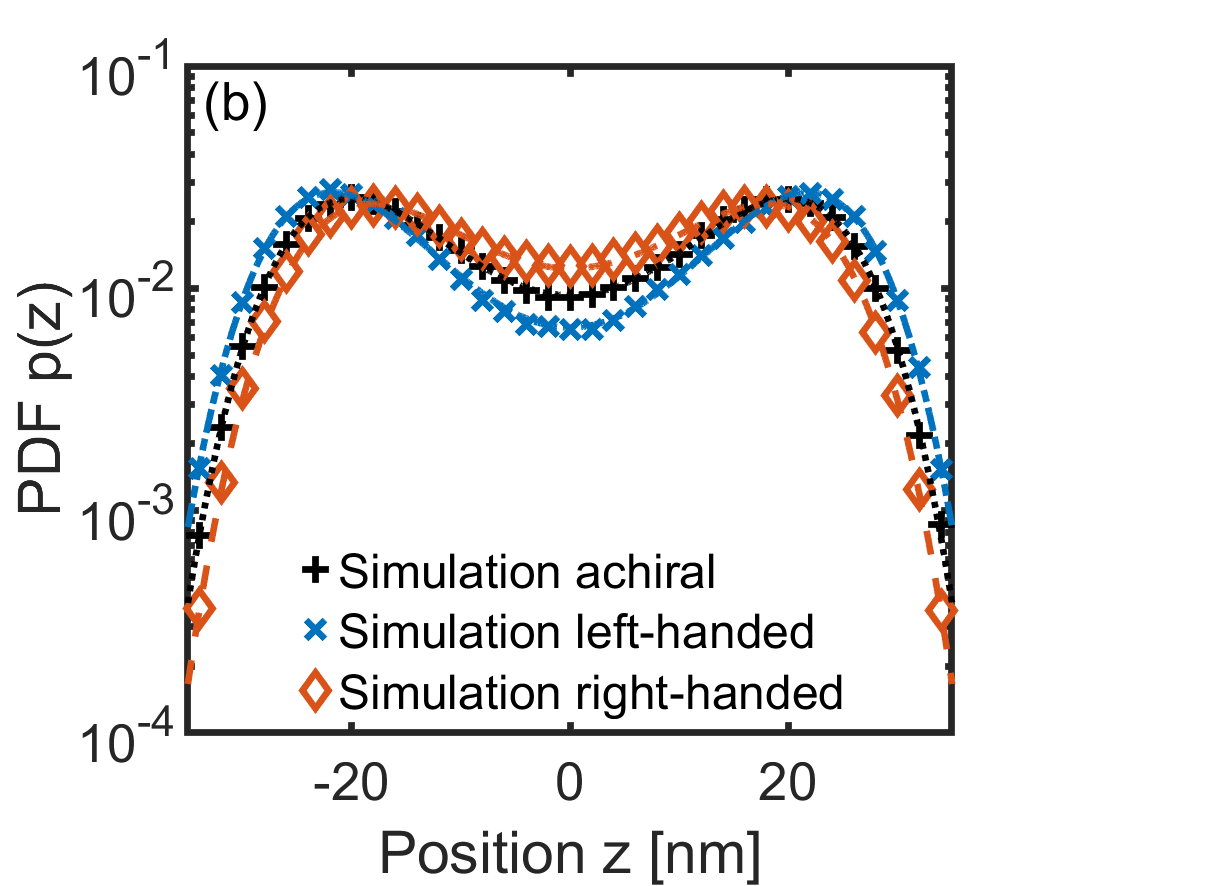}
    }
  \caption{(a) With $h_+=h_-$ and the polarization settings of the two beams detailed Fig. \ref{Fig:AchiralPotential}, the chiral coupling is reactive, leading to conservative chiral optical forces exerted on the diffusion of the chiral dipolar nanosphere ($\chi/\alpha=5\%$) within the bistable optical trap. Because these forces derive from a chiral potential $U_{\chi}(q,z)$, they combine with the achiral electromagnetic potential to form the potential energy surface $U_{\rm pot}(q,z)$. The resulting potentials at $q=0$  are plotted for the two opposite enantiomers: the ``right-handed'' (${\rm Re}\left[\chi\right] > 0$ in red) and the ``left-handed'' one (${\rm Re}\left[\chi\right] < 0$ in blue). The same is plotted in black for an achiral environment. The same differences would be induced for one fixed choice of enantiomer but using two opposite enantiomorphs for the reactive chiral optical field. (b) The corresponding normalized probability density functions (PDFs) are evaluated from our one-dimensional model (lines) and three-dimensional simulations (symbols) as detailed in Appendix \ref{Appendix:Simu}. Although the achiral and chiral PDFs appear only slightly different, their enantioselective character is manifest, revealing a chiral discriminating thermodynamics.  }
  \label{Fig:PDFReac}
\end{figure}

The second important thermodynamic consequence is the enantioselective character of the free energy difference $\Delta \mathcal{F}_{\chi}$ considering that ${\rm Re}\left[\chi\right]$ has opposite signs for different enantiomers of the chiral dipole and that $K(q,z)$, being a pseudoscalar, changes sign for the enantiomorphs (parity operation) of the chiral optical standing-wave. Chiral coupling therefore has the capacity to yield a new potential energy surface that depends on both the chirality of the dipole and of the optical field. This dual enantiomeric and enantiomorphic dependence of $W_\chi^{\rm cons}$ is the manifestation of a truly chiral discriminating thermodynamic process, concentrating one enantiomer towards the center and the other towards the outside of the double well as illustrated Fig. \ref{Fig:PDFReac}. 

Fig. \ref{Fig:PDFReac} (a) indeed displays the initial optical potential energy and the changes induced on it by the chiral density $K(\myvec{r})$ through the chiral coupling. As seen in panel (a), the contribution of the chiral potential, proportional to $W_{trap}(\myvec{r})$ with a sign determined by the enantiomeric form of the dipole, either enhances the trapping component of $W_E(\myvec{r})$ for ``right-handed'' eniantomers (${\rm Re}\left[\chi\right] >0$) or favors its interferential component $W_{inter}(\myvec{r})$ for ``left-handed'' ones (${\rm Re}\left[\chi\right] <0$). 

In the steady-state regime, that implies the detailed balance $\widetilde{J}_z^+=\widetilde{J}_z^-$, we also plot in panel (b) the probability density function (PDF) evaluated on the optical axis at $q=0$ which is simply given by
\begin{eqnarray}  
p_{\chi}^{\rm reac}(0,z)=C\cdot e^{-\frac{U_{\rm opt}(0,z)+U_{\chi}(0,z)}{k_{\rm B}T}},
\end{eqnarray} 
with $U_{\rm opt}(0,z)=-{\rm Re}\left[\alpha\right] W_E(0,z)$, $U_{\chi}(0,z)=-{\rm Re}\left[\chi\right] K(0,z) / \omega\sqrt{\varepsilon_f\mu_f}$ and $C$ a normalization factor evaluated such that $\int_{-\infty}^{+\infty} \dd{z} p_{\chi}^{\rm reac}(0,z) = 1$.

\section{Steady-state in the dissipative chiral coupling}  \label{sec:diss}

If we change the polarizations to $h_+=-h_-$, the standing-wave now carries a chiral flux with no chiral density. As a consequence, a dissipative chiral force is exerted on the diffusing chiral dipole. As we explain below, this mere change of polarization that switches the chiral optical environment from reactive to dissipative leads to a totally different thermodynamics. 

\begin{figure}[htb!]
  \centering{
    \includegraphics[width=0.8\linewidth]{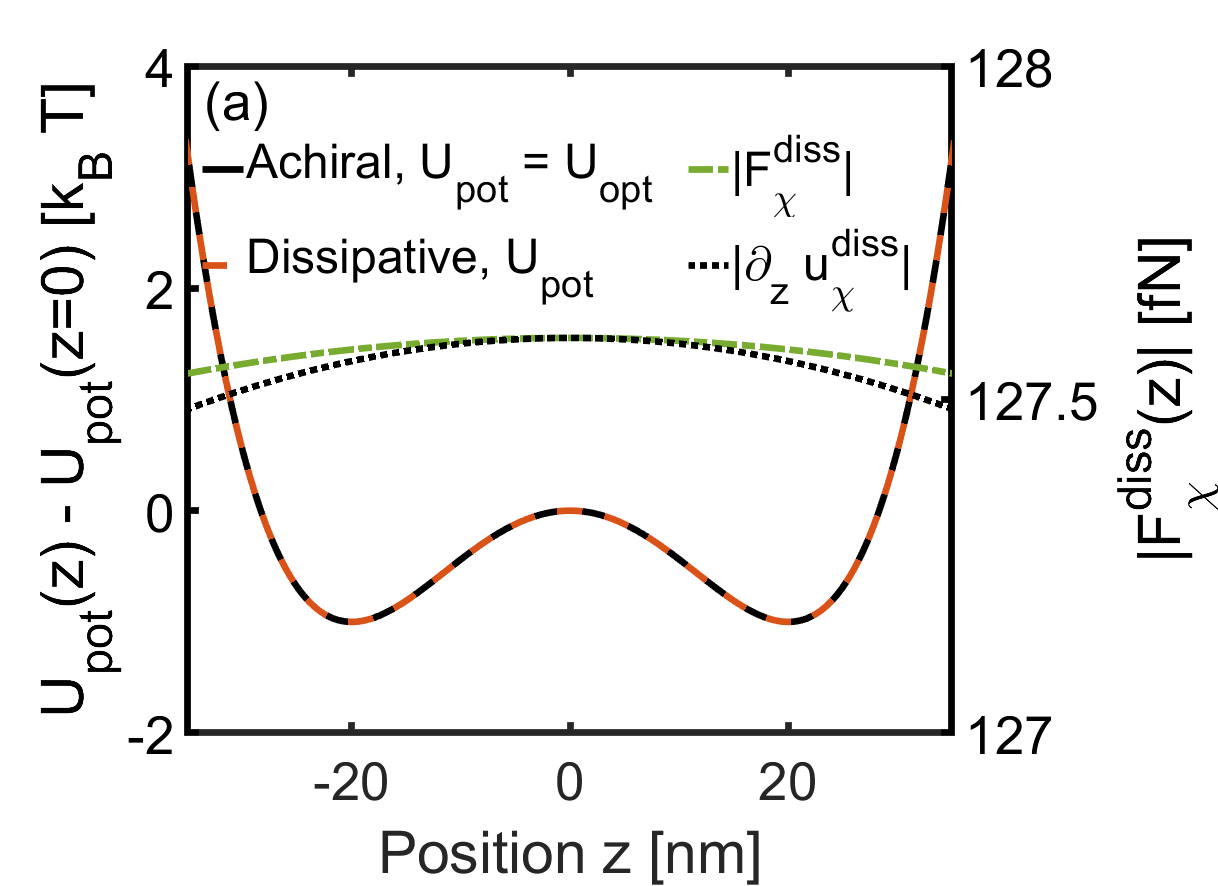}
    \includegraphics[width=0.8\linewidth]{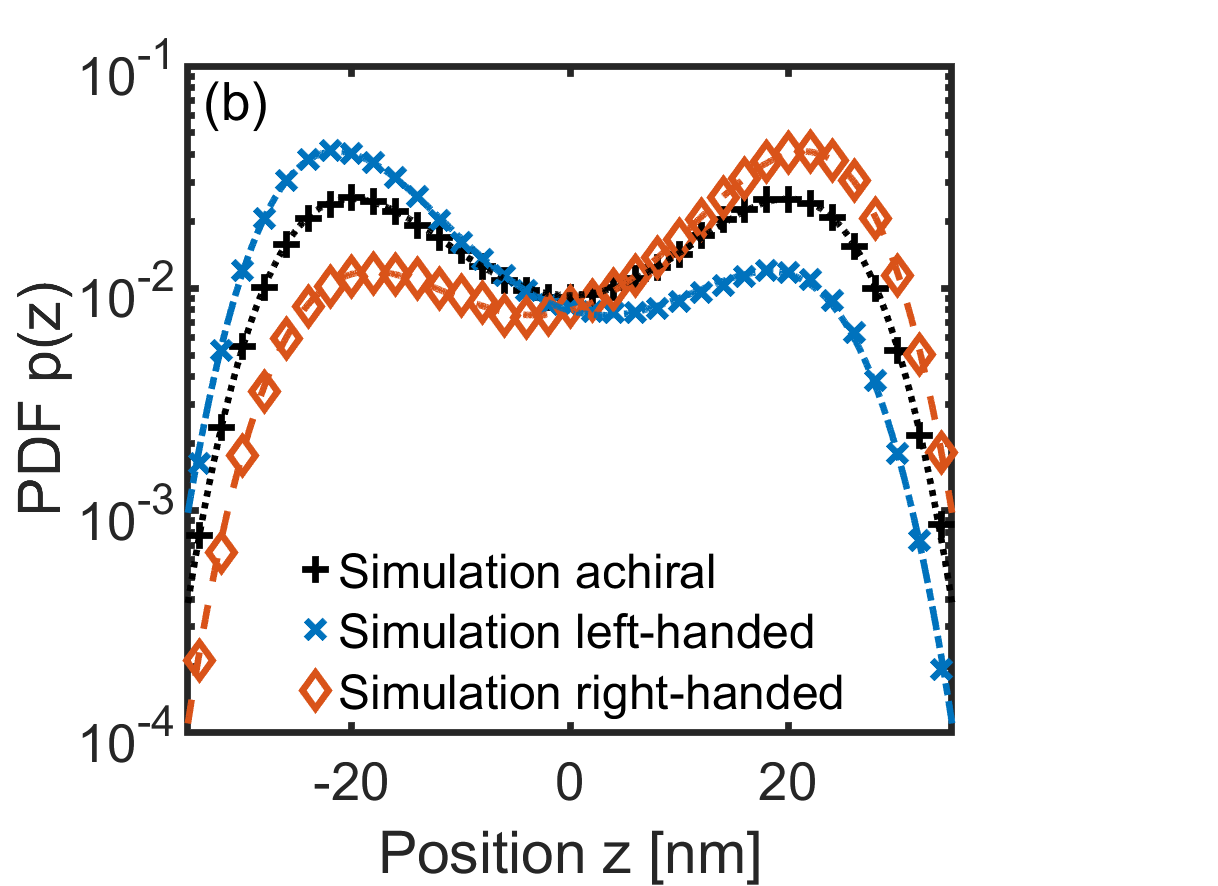}
    }
  \caption{(a) A dissipative chiral coupling is induced the chiral nanosphere ($\chi/\alpha=5\%$) for $h_+=-h_-$ in the two beams and the polarization settings detailed Fig. \ref{Fig:AchiralPotential}. The angle formed between the polarization axes of the beams and their helicities are fixed so as to lead to the generation of a chiral flux $\myvec{{\bf \Phi}}(q,z)$ -and zero chiral density- with the same achiral optical potential energy density $U_{\rm opt}(q,z)$ as for an achiral environment (in black, same as in Fig. \ref{Fig:AchiralPotential} (a)) again calculated at $q=0$. Our chiral dipole now couples to the chiral optical environment through ${\rm Im}\left[\chi\right]$ with chiral dissipative forces $F_{\chi}^{diss}(q,z)$ that are opposed for opposite signs in ${\rm Im}\left[\chi\right]$ -the same sign inversion appears if, instead of changing dipole enantiomers, one changes electromagnetic field enantiomorphs. Note that in the paraxial approximation of the model with $\Delta\ell\ll z_R$, we have $\Delta F_\chi^{\rm diss}(q,z)\ll F_\chi^{\rm diss}(q,z)$ -see main text. Here, we plot the profile of a second order development of $F_\chi^{\rm diss}(q,z)$ -black line, with the associated scale on the right-hand side of the graph. Because the dissipative chiral force is non-conservative, it does not contribute to the potential free energy surface as it is the case for the reactive chiral coupling. (b) The influence of the chiral dissipative force is seen on the modified steady-state probability density function of the chiral dipole in the bistable trap, shifted in the direction of the force for both our one-dimensional model (lines) and three-dimensional simulations (symbols) -- details in Appendix \ref{Appendix:Simu}. This modified PDF reveals the strong chiral discriminating action of $F_{\chi}^{diss}(0,z)$ with respect to the two local maxima at $z_A$ and $z_C$. We give in the main text the thermodynamics interpretation of this result.  }
  \label{Fig:PDFDiss}
\end{figure}

Because the dissipative chiral force is non-conservative with $\myvec{\nabla}\times \myvec{F}_\chi^{\rm diss}(q,z)\neq \myvec{0}$, it is not possible to derive it from a chiral potential, as it was the case for the reactive chiral coupling. But despite the non-conservative nature of the chiral force, we will solve the steady-state Fokker-Planck equation 
\begin{eqnarray}
\widehat{j}^+_z(q)&=&-\frac{1}{\gamma}\left(\partial_zU_{\rm opt}(q,z)-{\bf F}^{\rm diss}_{\chi}(q,z)\cdot\myvec{z}\right)\widehat{p}^+(q,z) \nonumber  \\
&&-D\partial_z \widehat{p}^+(q,z) \label{fokk:diss}
\end{eqnarray} 
analytically by making use of the fact that under the paraxial approximation, the $z$-dependence of the chiral flux $\myvec{\Phi}(q,z)$ is very slow over the distance $\Delta z \sim\Delta\ell$ separating the two local minima. The projected chiral dissipative force $F_\chi^{\rm diss}(q,z)=\hat{\myvec{z}}\cdot \myvec{F}_\chi^{\rm diss}(q,z)$ is thus such that
\begin{eqnarray}
\Delta F_\chi^{\rm diss}(q,\Delta z)= \partial_{z^2}^2 F_\chi^{\rm diss}(q,0)\Delta z^2 \ll F_\chi^{\rm diss}(q,0)   \label{eq:slow}
\end{eqnarray}
given that the symmetry of the force field imposes $\partial_{z} F_\chi^{\rm diss}(q,0) = 0$. This is well seen in Fig. \ref{Fig:PDFDiss} (a) where over $\Delta z \approx\SI{100}{\nano\meter}$, $|F_\chi^{\rm diss}(0,\Delta z)-F_\chi^{\rm diss}(0, 0)|/|F_\chi^{\rm diss}(0, 0)| = \Delta z^2/z_R^2 \approx 3.4\times 10^{-3}$. 

This slow-varying evolution of $F_\chi^{\rm diss}(q,z)$ throughout the bistable region makes it possible to approximate the dissipative force by its Taylor expansion around $(q, z=0)$ in $z$. Given the parity of the force, only pair orders are present and coefficients evolve in $1/(z_R^{2n})$ where $n$ is the expansion order in $z$. Choosing an arbitrary expansion order, we can tune the precision of the approximation of the chiral force field over a given volume inside the trap. This expansion can then be integrated as a pseudo-potential $u_{\chi}^{\rm diss}(q,z)$. Note that as stressed already above, it is not strictly possible to find a chiral potential for the dissipative chiral force. Our approximation here therefore neglects the fact that the pseudo potential defined is dependent on $q$ while there is no associated radial force. In an effective way, we use a pseudo-one dimensional model with a radial $q$ parameter, exploiting the fact that in one dimension, all forces can be derived from a potential. For the sake of simplicity, we will use a second order development for our model for defining $F_\chi^{\rm diss}(q,z) =-\partial_z u_{\chi}^{\rm diss}(q,z)$. This pseudo-potential approach will help us solving the steady-state Fokker-Planck equation (\ref{fokk:diss}) using the same steepest-descent approach and the $q-$parity of $\myvec{\Phi}(q,z)$. We can then evaluate analytically the probability density function under dissipative chiral coupling plotted in Fig. \ref{Fig:PDFDiss} (b) --see below. 

Under such an approximation, the Fokker-Planck equation (\ref{fokk:diss}) can be directly integrated, leading to escape rates modified by the external chiral dissipative force field as 
\begin{eqnarray}  
\widehat{\kappa}_{A\rightarrow C} &=& \frac{\widehat{J}^+_z}{\widehat{n}^+_A}\nonumber  \\
&&\simeq  \kappa_{A\rightarrow C} \times e^{+\frac{F_\chi^{\rm diss}(0,0)  \left(z_B-z_A\right)}{k_{\rm B} T}} , \label{rates:diss1}\\  
\widehat{\kappa}_{C\rightarrow A} &=& \frac{\widehat{J}^-_z}{\widehat{n}^-_C}\nonumber  \\
&&\simeq  \kappa_{C\rightarrow A}  \times e^{+\frac{F_\chi^{\rm diss}(0,0)  \left(z_B-z_C\right)}{k_{\rm B} T}}, \label{rates:diss2}
\end{eqnarray} 
with $\widehat{a}\simeq a$, $\widehat{b}\simeq b$ and $\widehat{c}\simeq c$ as we verified here too.

Because the chiral electromagnetic fields continuously transfer, through dissipation, mechanical energy to the chiral dipole immersed in this dissipative chiral environment, our system behaves as a nonequilibrium steady-state system where the chirality of the probe becomes a thermodynamic parameter. The thermodynamic consequence of the emergence of a dissipative chiral optical force is a bias put on the probability distribution function of positions from both sides of the waist. In this dissipative coupling, the PDF is evaluated in the stationary regime, on the optical axis, using a nonequilibrium potential $p^{\rm diss}_{\chi}(0,z)=C\exp\left[-\varphi(0,z)\right]$ where we have, within the pseudo-potential $u_{\chi}^{\rm diss}(0,z)$ approach, $\varphi(0,z)=(U_{\rm opt}(0,z)+u_{\chi}^{\rm diss}(0,z))/k_{\rm B}T$, and the normalization $C^{-1}=\int_{-\infty}^{+\infty}\dd z \exp{-\varphi(0,z)}$ \cite{SekimotoBook,Seifert2010,Speck2012}. It is plotted in Fig. \ref{Fig:PDFDiss} (b).

As already emphasized, the chiral coupling intertwines the chirality of the dipole with the chirality of the field, while leaving untouched the achiral bistable potential. For this reason, the bias depends on both the enantiomeric form of the dipole via ${\rm Im}\left[\chi\right]$ and the enantiomorphic form of the field through the chiral flux ${\bf \Phi}(q,z)$. But contrasting with the reactive case, the dissipative chiral action cannot be framed into a chiral contribution to the potential energy landscape. In such an ``exclusive'' framework indeed, the chiral force contributes to the thermodynamics as a dissipative work $F_\chi^{\rm diss}(q,z)\delta z=\delta W_\chi^{\rm diss}$ and not as a free energy change \cite{jarzynski2007comparison}. This fundamental difference in the thermodynamics between the reactive and the dissipative chiral couplings has important consequences as we now see.

The chiral dissipative force break the symmetry of the escape rates 
\begin{eqnarray}  
\frac{\widehat{\kappa}_{A\rightarrow C}}{\widehat{\kappa}_{C\rightarrow A}} = e^{+\frac{F_\chi^{\rm diss}(0, 0)  \left(z_C-z_A\right)}{k_{\rm B} T}}.  \label{rates:diss}
\end{eqnarray} 
and act as a chiral source of heat $\Delta Q_\chi =Q_\chi^{C\rightarrow A}-Q_\chi^{A\rightarrow C}=F_\chi^{\rm diss}(0, 0)  \Delta\ell / \left(k_{\rm B} T\right)$ transferred to the surrounding fluid in the trap. We note that the sign of this transfer is determined by the orientation of the chiral force with respect to oriented inter-well distance $\Delta\ell>0$, in other words depends, via ${\rm Im}\left[\chi\right]\lessgtr 0$, on the enantiomeric form of the dipole. This enantiodependence is observed in Fig. \ref{Fig:PDFDiss} (b) in the difference in the probability density function between the two wells.   

The heat transfer can be described as an associated entropy production $\Delta S_\chi=\Delta Q_\chi / T$ during the diffusion of the dipole from one well to the other. This production of entropy is only related to the dissipative chiral dynamics and can be associated with the ``entropy penalty'' expected for any deracemization process, as mentioned in the Introduction \cite{amabilino2011spontaneous,palmans2017deracemisations}. 

\section{Stochastic simulations: trajectories and probability density functions}  \label{sec:simuEns}

Once the model exposed and analytically solved, we can go a step further by simulating the three-dimensional instantaneous motion $\myvec{r}(t)$ of the chiral dipole inside the optical trap when the polarizations are set to induce chiral optical environments. To do so, we solve the overdamped Langevin equation 
\begin{eqnarray}
\gamma \ d_t \myvec{r} = - \myvec{\nabla} U_{\rm opt}(\myvec{r})+\myvec{F}_\chi(\myvec{r})+\myvec{F}_{\rm th}(t)
\end{eqnarray}
in the achiral bistable optical potential $U_{\rm opt}(\myvec{r})$ with $\myvec{F}_{\rm th}(t)$ the thermal random force of zero mean that satisfy the fluctuation-dissipation theorem. We include in this Langevin equation the three-dimensional chiral force fields (axial and radial) whose expressions have been reminded in Section \ref{remind}. In Fig. \ref{Fig:AchiralPotential} (a), the electromagnetic field intensity was adjusted so that the bistable barrier separating the two local potential minima is set to a height of one $k_{\rm B}T$. 

The simulations are performed in achiral $(\myvec{F}_\chi(\myvec{r})={\bf 0})$, chiral reactive $(\myvec{F}_\chi(\myvec{r})=\myvec{F}^{\rm reac}_\chi(\myvec{r}))$ and chiral dissipative $(\myvec{F}_\chi(\myvec{r})=\myvec{F}^{\rm diss}_\chi(\myvec{r}))$ configurations, using the same polarization settings as those involved in Fig. \ref{Fig:AchiralPotential}. Again, the chirality of the trapped nanosphere is set to $\chi/\alpha = 5\%$. Simulations are run for a racemic mixture of chiral dipoles corresponding to $10^{4}$ trajectories per eniantomer in parallel, starting from an initial distribution of positions determined from the three-dimensional stationary probability density distributions evaluated by our model, for $5\times 10^{4}$ time steps. Simulation algorithms and methods are detailed in Appendix $\cdots$.

\begin{figure}[htb!]
  \centering{
    \includegraphics[width=0.7\linewidth]{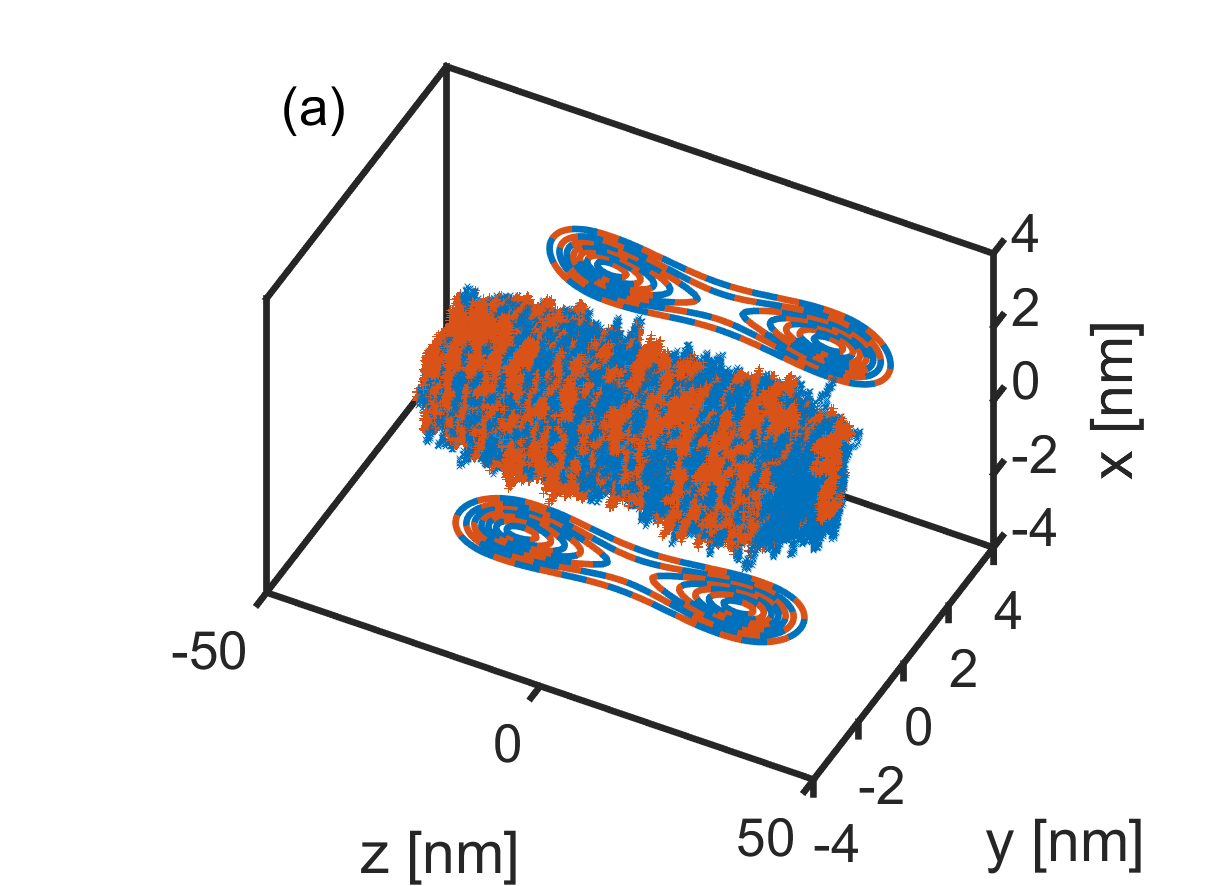}
    \includegraphics[width=0.7\linewidth]{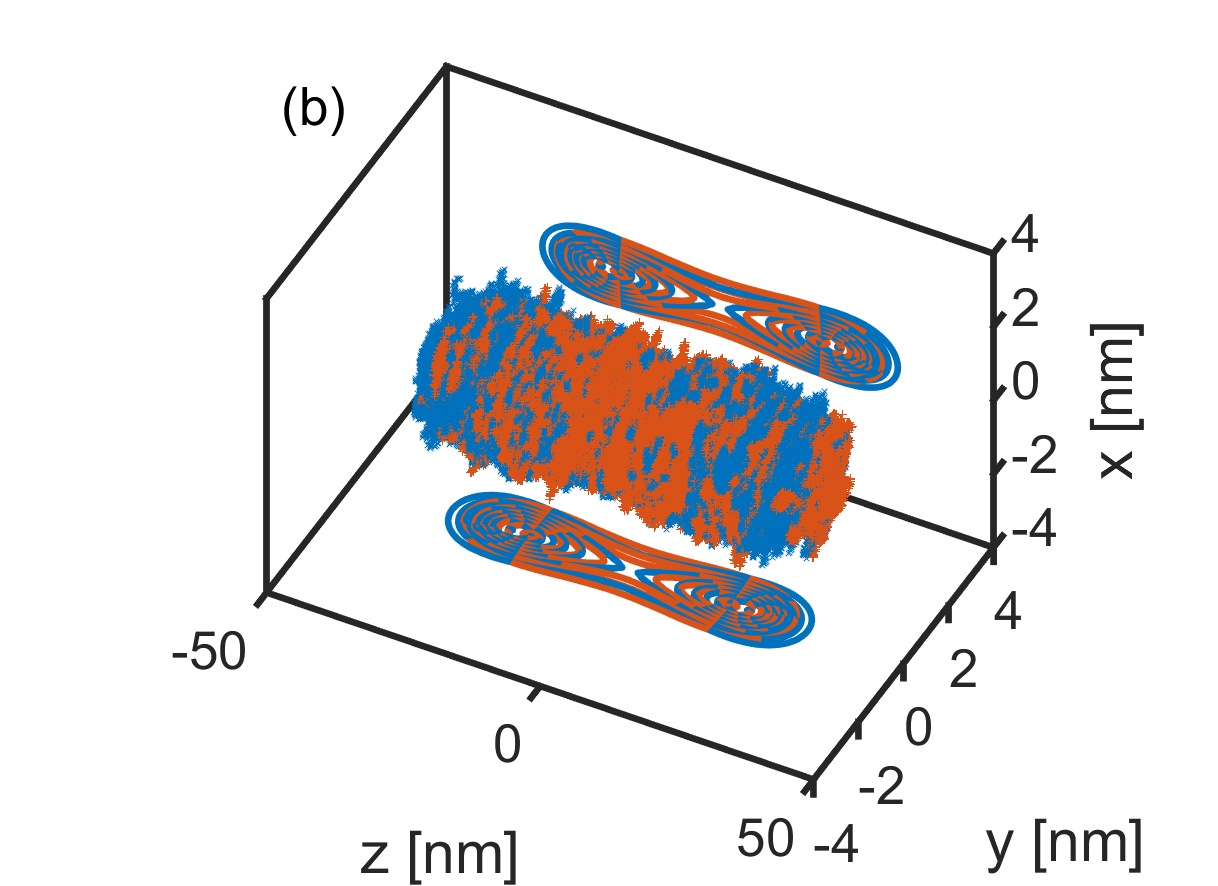}
    \includegraphics[width=0.7\linewidth]{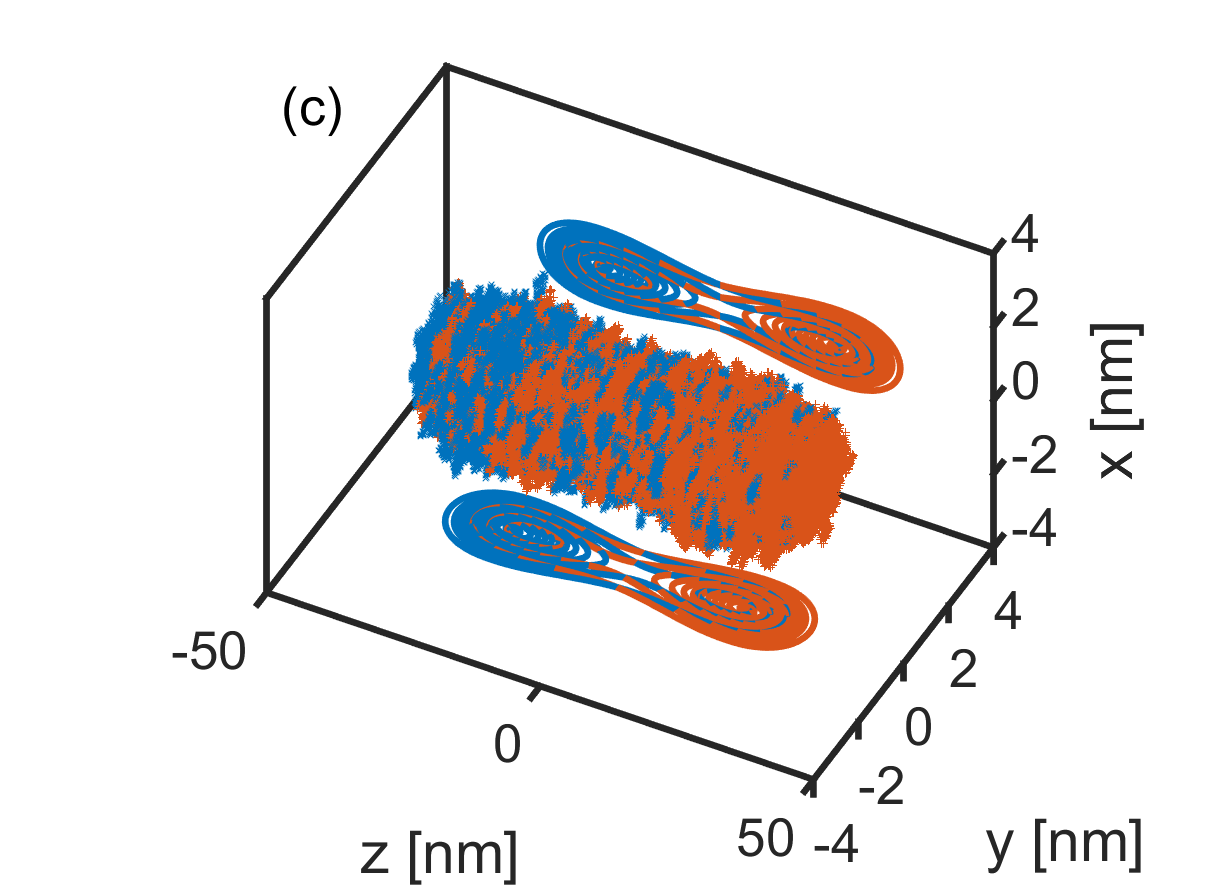}
    }
  \caption{Time-dependent positions of $R=\SI{20}{\nano\meter}$ Au nanospheres in a racemic mixture of \num{100} ``left-handed'' (in blue) and ``right-handed'' (in red) enantiomers, simulated for $\num{50000}$ time steps of $\SI{95.4}{\pico\second}$ for a total of $\SI{4.77}{\micro\second}$. These trajectories are randomly picked among the \num{1e4} trajectories used to form the simulated PDF displayed in Fig. \ref{Fig:PDFReac} (b) and \ref{Fig:PDFDiss} (b). We vary the chiral nature of the optical environment going through (a) an achiral optical environment, (b) a reactive chiral optical environment and (c) a dissipative chiral optical environment. We can see that in the achiral case, the distribution is totally independent from the enantiomeric form of the nanoparticle, as expected. In the reactive case in contrast, the chiral nanoparticles are either more concentrated towards the center of the optical trap for the family of ``right-handed'' enantiomers (${\rm Re}\left[\chi\right] >0$) or moved away to external regions for the ``left-handed'' enantiomers (${\rm Re}\left[\chi\right] <0$). In the dissipative case, the chiral coupling is capable of inducing a deracemization process by progressively localizing enantiomers to different wells in strict relation with their enantiomeric ${\rm Im}\left[\chi\right]\lessgtr 0$ forms. For each of the different types of couplings, the contour plots shown on the axis planes correspond to the predictions of the $p_{\rm opt}(q,z),p_{\chi}^{\rm reac}(q,z),p_{\chi}^{\rm diss}(q,z)$ PDFs given by our one-dimensional model. As well seen, our model successfully reproduces both the axial and radial distributions.}
  \label{Fig:3DTraj}
\end{figure}

Within all the available states that lie below the level set by the temperature and the simulation time, these results perfectly reveal how the Brownian motion probes the chiral optical environment, where the chiral coupling bias the diffusion driven by thermal fluctuations. The spatial distributions of positions numerically calculated and shown in Fig. \ref{Fig:3DTraj} clearly reveal this bias. In the achiral case of Fig. \ref{Fig:3DTraj} (a), the distribution does not depend on the enantiomer while both reactive -Fig. \ref{Fig:3DTraj} (b)- and dissipative -Fig. \ref{Fig:3DTraj} (c)- cases are enantiodependent. With the chosen optical enantiomorph, we see in the reactive case that an optically trapped ``right-handed'' enantiomer with ${\rm Re}\left[\chi\right] >0$ is more concentrated towards the trapping maximum than for the opposite ${\rm Re}\left[\chi\right] <0$ enantiomer. In the dissipative coupling, the enantiomers are clearly spatially separated. These signatures seen on trajectories complement Sec. \ref{sec:diss} in the demonstration and characerization of a genuine optomechanical deracemization process. 

For each three simulations, we can also build the corresponding PDF and compare them with the one-dimensional ones evaluated in our model. Although it would be possible to evaluate, in each simulation, the axial distribution using only positions of the nanospheres that lie very close to the axis, it turns more favorable in terms of statistics to integrate this distribution in each axial plane, leading to the distribution
\begin{eqnarray}
p_{sim}(z) &=& \frac{\int\limits_0^{+\infty}{\rho}(q, z) 2 \pi q \dd{q}}{\int\limits_{-\infty}^{+\infty}\int\limits_0^{+\infty}{\rho}(q, z) 2 \pi q \dd{q}\dd{z}}
\end{eqnarray} 
where $\rho(q, z)$ is the volume density of particles positions given by the simulations. This density is linked to the actual distribution of all simulated positions $p(q, z)$ and to the number $N$ of position data by the relation $\rho(q, z) = N p(q, z)$. Under the $q$-parity hypothesis of Section \ref{sec:reac}, this distribution is expected to match the predictions of our one-dimensional model.

In Fig. \ref{Fig:PDFReac} (b) and Fig. \ref{Fig:PDFDiss} (b), we compare the simulated PDFs to the model axial PDFs, respectively $p_{\chi}^{\rm reac}(0,z)$ and $p_{\chi}^{\rm diss}(0,z)$. The excellent agreement validates in the specific case of dissipative chiral coupling, the pseudo-potential approach of our model. It also validates for all cases that one can use the corresponding stationary PDF of the model for initializing the simulations, as discussed in Appendix \ref{Appendix:Simu}. 

\section{Stochastic simulations: well residency times statistics}  \label{sec:simuSin}

We now look at single long diffusive trajectories of one chiral nanosphere within the optical trap, thermally activated from one local well to the other. For such a study, it is important to have good statistics on barrier-crossing events and we therefore chose to use here a number of trajectories reduced in comparison with the ensemble simulations of Sec. \ref{sec:simuEns} but allowing to calculate over longer times -over $2.9$ ms corresponding to $3\times 10^6$ points with a time step $dt=0.95$ ns. 

\begin{figure*}[htb!]
  \centering{
    	\includegraphics[width=0.8\linewidth]{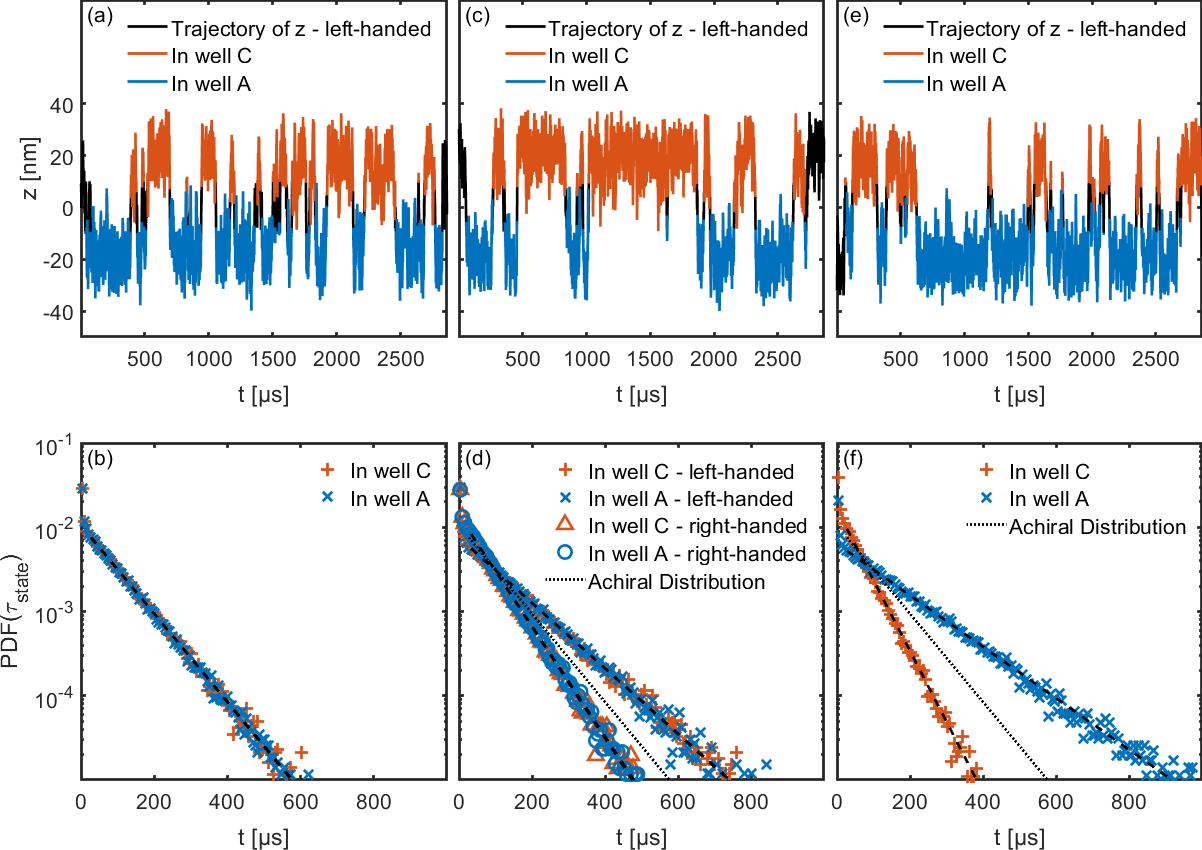}
    }
  \caption{(a) One trajectory simulated over $\SI{2.9}{\milli\second}$ with time steps of $\SI{0.95}{\nano\second}$ corresponding to the diffusion of an optically trapped nanosphere in the achiral bistable optical potential energy landscape displayed in Fig \ref{Fig:AchiralPotential} (b). Jumps are identified using a hysteresis of $\sigma=\SI{10}{\nano\meter}$. The trapped periods are highlighted for well A in blue and C in red. The first and last events are excluded as explained in Appendix \ref{Appendix:Jump_PDF}. (b) Cumulated statistics (over \num{4096} trajectories) of the residency time in both wells $A$ ($z<0$, in blue) and $C$ ($z>0$, in red) for the symmetric bistable potential. For times larger than the relaxation time of the well, the statistics follow a Poisson distribution, as expected from Kramers' theory. The average residency time in one well corresponds to the inverse of the linear slope of the distribution associated to that well, plotted in a lin-log scale. (c) One trajectory simulated within a chiral reactive environment for a ``left-handed'' chiral nanosphere ($\Re[\chi]<0$). As can be seen, the nanosphere spends an equivalent time in both wells, but its jumps frequency is slowed down compared to the achiral case. As seen in (d), the same phenomenon appears with the same degeneracy but with an acceleration for a ``right-handed'' nanosphere. (e) One trajectory, this time simulated within a chiral dissipative environment for a ``left-handed'' chiral nanosphere (${\rm Im}\left[\chi\right]>0$). As can be seen, the chiral nanosphere does not spend an even time between the two wells, in stark contrast with the achiral bistable case and the chiral reactive coupling case. The breaking of symmetry when the dissipative chiral coupling is switched on is reflected in the splitted exponential laws associated with the two wells calculated in (f).}
  \label{Fig:Jumps}
\end{figure*}

Fig. \ref{Fig:Jumps} (a) shows a time trace $z(t)$ along the optical $q=~0$ axis of one such long diffusing trajectory in the achiral bistable potential defined in Sec. \ref{sec:bistable}. The time trace clearly reveals the stochastic motion of the trapped nanoparticle that ``jumps'' from one well to the other (ca. $40$ jumps for a $2.9$ ms trajectory). Such jumps are described by a Poisson statistics where the residency time $\tau_{i}$ in each well $i=A,C$ follows an exponential law $P(\tau_i)=\exp(-\tau_i/\langle\tau_i\rangle)/\langle\tau_i\rangle$, where $\langle\tau_i\rangle$ is the mean residency time in well $A$ or $C$  \cite{simon1992escape}. As explained in Appendix \ref{Appendix:Jump_PDF}, the evaluation of such distributions demands a careful identification of the jumps, accounting for the possible re-crossing events present in all thermally activated barrier-crossing diffusive systems. Fig. \ref{Fig:Jumps} (b) shows that the symmetry of the achiral bistable potential leads to identical exponential laws for the residency times inside each well $A$ and $C$. 

A time trace $z(t)$ along the optical $q=~0$ axis simulated in the case of a reactive chiral coupling is displayed in Fig. \ref{Fig:Jumps} (c). The degeneracy in residency times preserved in the reactive chiral coupling is clearly observed in Fig. \ref{Fig:Jumps} (d), with a difference in the residency times corresponding to the differences in the well depths shown in Fig. \ref{Fig:PDFReac}. The discriminative action of the chiral reactive coupling is measured here on the exponent differences between the two enantiomer families.  

In the case of dissipative chiral coupling in contrast, we already know from Sec. \ref{sec:diss} that the degeneracy between the two wells is broken. This is perfectly seen on the time trace displayed in Fig. \ref{Fig:Jumps} (e) diffusion dynamics of the chiral nanosphere, and more clearly on the probability distributions of the residency times in Fig. \ref{Fig:Jumps} (f).  The observed tendency to spend more time in well $A$ than in well $C$ for a ``left-handed'' enantiomer (${\rm Im}\left[\chi\right]>0$) is in agreement with the probability density functions plotted in Fig. \ref{Fig:PDFDiss} (b). This symmetry breaking is responsible for the deracemization process observed in Fig. \ref{Fig:3DTraj} (c) above when dealing with a pure racemic mixture composed of a large, even number of optically trapped ``left-'' and ``right-handed'' enantiomers. In this dissipative case, we measure from the exponential laws a ratio between the residency times $\langle\tau_{C}\rangle /\langle\tau_{A}\rangle ={0.36}$. 

This measured ratio can be directly compared to the model with $\langle\tau_{C}\rangle /\langle\tau_{A}\rangle =\kappa_{A\rightarrow C} /\kappa_{C\rightarrow A}$. The ratio is evaluated using Eq. (\ref{rates:diss}), reaching $\kappa_{A\rightarrow C} /\kappa_{C\rightarrow A}|_{\rm model}=0.29 $, which clearly departs from the simulated result. This disagreement suggests that the role of the force field contributions cannot simply be limited to quadratic approximations taken at the wells' minima, as it is done in the steepest descent approach of our one-dimensional model. Accounting for higher order terms, i.e. the anharmonicity of the wells around the barrier, is required for quantitative comparisons.

This is confirmed if we now look at three-dimensional PDFs. This ratio indeed can also be related to the populations inside each well according to $\langle\tau_{C}\rangle /\langle\tau_{A}\rangle = n_C^- J_z^+/ n_A^+ J_z^-$ and therefore can be directly evaluated within the detailed balance, $J_z^+=J_z^-$ using probability densities. In this approach, we extend the expression of the detailed balance stationary PDF given in Sec. \ref{sec:diss} to three-dimensions with $p(q,z)=C\exp \left[-\varphi(q,z)\right]$, with $\varphi(q,z)=(U_{\rm opt}(q,z)+u_{\chi}^{\rm diss}(q,z))/k_{\rm B}T$. The population in one well is then evaluated by an integration of the PDF restricted over the well and therefore 
\begin{eqnarray}
\frac{n_C^-}{ n_A^+} = \frac{\int\limits_0^{\infty}\dd{q}2\pi q\int\limits_{0}^{\infty}\dd{z} p(q,z)}{ \int\limits_0^{\infty}\dd{q}2\pi q\int\limits_{-\infty}^{0}\dd{z} p(q,z)} =0.36.\nonumber  
\end{eqnarray} 
The perfect agreement with the simulations confirms the (expected) quantitative importance of accounting for the anharmonic curvature of the potential generated by the interfering Gaussian beams.

The important scope of these results is to show that it is possible to detect and measure the presence of chiral optical forces by looking at the average of the residency times of each of the wells rather than at the forces themselves. Considering that these residency times are exponentially sensitive to either the chiral free energy (in the case of reactive chiral coupling) or the chiral heat (in the case of dissipative chiral coupling), one expects such an approach to yield a high resolution in detection of chiral optical forces and in the resolution of the chiral discriminative thermodynamics at play when the chiral coupling is switched on. 

In particular, the escape rates $\kappa = 1/\langle\tau\rangle$ corresponding to each three optical landscapes can be extracted from the measurements of the average residency times using the Poisson statistics. Since we have shown that our pseudo-potential approach allows to predict very precisely the distribution of positions and since the optical landscapes can, each, be set very precisely, it thus becomes possible to perform an \textit{absolute} determination of $(i)$ ${\rm Re}\left[\chi\right]$ by measuring $\langle\tau^{\rm achiral}\rangle / \langle\tau_{A,C}^{\rm reac}\rangle$ --see Eqs. (\ref{rates:reac1},\ref{rates:reac2})-- and of $(ii)$ ${\rm Im}\left[\chi\right]$ by measuring $\langle\tau^{\rm achiral}\rangle / \langle\tau_{A,C}^{\rm diss}\rangle$ --see Eqs. (\ref{rates:diss1},\ref{rates:diss2}). This determination is done at the single nanoparticle level, and as such draws promising detection capacities in the context of artificial chiral matter engineering at the nanoscale \cite{schnoering2018three,vinegrad2018determination,spaeth2019circular,sachs2020}. 

\section{Conclusion}
\label{sec:conclusion}

We studied, in the framework of the Fokker-Planck equation, the stochastic motion of an overdamped Brownian chiral probe optically trapped, diffusing in a bistable potential energy landscape formed in the standing-wave of two counter-propagating Gaussian beams. We analyzed in this framework the modifications of the escape rates when a chiral coupling is induced between the probe and the optical field. We summarize the main results:
\begin{itemize}
\item the chiral coupling mediated by optical forces can be switched on inside the optical trap simply by selecting the polarizations of the counter-propagating beams forming the initial, achiral, bistable potential, while keeping fixed the energy densities,
\item chiral coupling (of reactive and/or dissipative nature) leads to modifications of the thermodynamics of the thermal activation of the barrier that are enantiospecific and dependant on the enantiomorphic configurations of the chiral optical environment,
\item  more precisely, reactive coupling takes the form of conservative chiral optical forces and thus contributes as an additional free energy term to the potential energy of the bistable trap. The modifications of the free energy landscape either strengthen the trapping potential or decreases the barrier height of the chiraly-dressed potential. These modifications can be swapped by changing the enantiomer within a fixed chirality of the optical environment, or the optical enantiomorph for a chosen nanoparticle enantiomer,
\item the dissipative coupling yields non-conservative chiral forces that ``exclusively'' work in the thermal activation thermodynamics. In this nonequilibrium steady-state of the system, the dissipation of heat to the thermal bath is responsible for lifting the degeneracy of the probability density function between the two local minima of the bistable potential. This breaking of the initial mirror symmetry of the bistable trap takes the form of an enantiospecific contribution to the thermodynamics,
\item the contribution of both types of coupling to the global thermodynamics is also observed at the level of stochastic simulations of the Langevin equation for trajectory ensembles in the presence of external chiral forces. The simulations clearly show in particular the chiral discriminatory nature of the dissipative coupling that constitutes an explicit example of a deracemization process analyzed from the thermodynamics viewpoint,
\item at the level of Langevin dynamics of single diffusing trajectories thermally activated over the barrier separating the two wells, the same results are reached by measuring the Poisson statistics of the residency times for each local minima of the bistable potential without, and with, chiral coupling. Measuring a difference in the average residency times in the case of the dissipative chiral coupling demonstrates, from the single trajectory viewpoint, the optomechanical deracemization process,
\item approaching the problem from the residency time point of view shows how one can probe the thermodynamics of the system from time measurement sequences only rather than from more demanding force measurements,
\item and how one can obtain an \textit{absolute} measurement of both the real and imaginary parts of the chiral polarizability of a single nanoparticle by extracting, from the Poisson statistics, the average residency times in the achiral, chiral reactive and chiral dissipative coupling schemes.
\end{itemize}

Overall, our results illustrate how the chiral coupling transforms chiral degrees of freedom into true thermodynamic control parameters. They open a rich playground to further exploring chiral light-matter interactions. The capacity of our model to solve the stochastic chiral bistable problem convinces us that the optical forces and residency times approaches can offer new and relevant insights on the thermodynamics of chiral systems immersed within chiral environments. Considering the ubiquity of such bistable landscapes in the realm of chirality, our model and our methods have a heuristic value that unfolds at the crossroad of chemistry and physics. In particular at the quantum level, further extending our results to chiral quantum optics \cite{lodahl2017chiral,mahmoodian2020} will give the possibility to study how chirality can impact quantum stochastic thermodynamics \cite{talkner2020,elouard2020}. This opens up new perspectives yet to be explored.

\section*{Acknowledgments}

We thank J. Crassous, J.-P. Dutasta, T. W. Ebbesen, M. W. Hosseini, and Ph. Lesot for discussions. This work was supported by the French National Research Agency (ANR) through the Programme d'Investissement d'Avenir under contract ANR-17-EURE-0024, the ANR Equipex Union (ANR-10-EQPX-52-01), the Labex NIE (ANR-11-LABX-0058 NIE) and CSC (ANR-10-LABX-0026 CSC) projects and the University of Strasbourg Institute for Advanced Study (USIAS) (ANR-10-IDEX-0002-02).

\begin{appendix}

\section{The dual-beam optical trap: optical landscapes and optical forces}
\label{Appendix:Force_model}

We extend here the simplified discussions of Secs. \ref{remind} and \ref{sec:bistable} in order to include magnetic force components and thus present the complete chiral force model in the dipolar regime \cite{canaguier2013mechanical}. We remind that our configuration consists in counter-propagating Gaussian beams identical in terms of intensity and spatial profile. In the paraxial approximation, this implies the cancellation of the Poynting vector $\myvec{\Pi}$, both its orbital and spin components. From a force viewpoint, this implies the absence of any radiation pressure force field. 

The polarization vectors $\myvec{e}_{\pm}$ associated with each beam can be described in a generic way with 
\begin{eqnarray}
\myvec{e}_{+} = (\sqrt{1-h_+}\myvec{e}_{l} + \sqrt{1+h_+}\myvec{e}_{r})/\sqrt{2}  \nonumber
\end{eqnarray}
for the beam propagating along the $z>0$ direction and 
\begin{eqnarray}
\myvec{e}_{-} = (\sqrt{1-h_-}e^{i(\delta-\delta\theta)}\myvec{e}_{l} + \sqrt{1+h_-}e^{i(\delta+\delta\theta)}\myvec{e}_{r})/\sqrt{2} \nonumber
\end{eqnarray}
for the counter-propagating beam ($z<0$ direction). In order to understand the role of the different polarization setting parameters, we stress that $h_+$ controls the helicity of the $z>0$ beam with $\myvec{e}_{+}$ varying from $\myvec{e}_{l}$ when $h_+ = -1$ to $(\myvec{e}_{l} + \myvec{e}_{r})/\sqrt{2}$ (i.e. linear state of polarization) when $h_+ = 0$, and to $\myvec{e}_{r}$ when $h_+ = +1$. The main polarization axis of the beam remains arbitrarily fixed and constitutes a degree of freedom for the axisymmetric system. The phase at time $t = 0$ is fixed as well, the system being invariant by translation of the initial time. For the counter-propagating beam, $\myvec{e}_{-}$ on the other hand, varies from $e^{i(\delta-\delta\theta)}\myvec{e}_{l}$ when $h_- = -1$ to $e^{i\delta} (e^{-i\delta\theta}\myvec{e}_{l} + e^{i\delta\theta}\myvec{e}_{r})/\sqrt{2}$ when $h_- = 0$, and to $e^{i(\delta+\delta\theta)}\myvec{e}_{r}$ when $h_- = +1$. As can be seen, the effect of the parameter $\delta$ is, in all cases, a global phase shift. It thus controls the relative phase between the counter-propagating beams. The $\delta\theta$ parameter, on the other hand, rotates the polarization axis, as can be most clearly seen by decomposing the circular polarization vectors in the linear polarization basis. These two parameters $\delta$ and $\delta\theta$ are expressed in radians, either as a phase angle, or as a physical angle between the main axis. 

Extending Sec. \ref{sec:bistable} to the magnetic case, we define the electric $W_E(\myvec{r})= \varepsilon_f \Esw\cdot\Esw^*/4$ and magnetic $W_H(\myvec{r}) =~\mu_f \Hsw\cdot\Hsw^*/4$ components to the time-averaged energy densities which can be expressed in terms of the trapping energy density $W_{trap}(\myvec{r})$ and an interference energy density $W_{inter}(\myvec{r})$ as: 
\begin{eqnarray}
W_E(\myvec{r}) &=& W_{trap}(\myvec{r}) + W_{inter}(\myvec{r}) \\
W_H(\myvec{r}) &=& W_{trap}(\myvec{r}) - W_{inter}(\myvec{r}) ,
\end{eqnarray}
where we have
\begin{eqnarray}
W_{trap}(\myvec{r}) &=& \frac{\mathcal{E}_0^2 w_0^2 \varepsilon_f}{2 w^2(z)} e^{-\frac{2 q^2}{w^2(z)}}\\
W_{inter}(\myvec{r}) &=& W_{trap}\left(\myvec{r}\right) \bigg(h_2 cos\left(\delta\theta\right) cos\left(\varphi(\myvec{r})\right) + \notag \\
& & h_1 sin\left(\delta\theta\right) sin\left(\varphi(\myvec{r})\right)\bigg) .
\end{eqnarray}

We also define the electric and magnetic ellipticities $\myvec{\Phi}_E(\myvec{r}) = i \omega \varepsilon_f \Esw\times\Esw^*/4$ and $\myvec{\Phi}_H(\myvec{r}) = i \omega \mu_f \Hsw\times\Hsw^*/4$ that can be summed to obtain the chiral flux introduced Section \ref{sec:chienv}, $\myvec{\Phi}(\myvec{r}) = \myvec{\Phi}_E(\myvec{r}) + \myvec{\Phi}_H(\myvec{r})$ \cite{canaguier2013mechanical,canaguier2015chiral}. Similarly to their scalar counterparts, these decompose into average $\myvec{\Phi}_{trap}(\myvec{r})$ and interference $\myvec{\Phi}_{inter}(\myvec{r})$ components. This time, however, both components depend on the polarization of the beams according to: 
\begin{eqnarray}
\myvec{\Phi}_E(\myvec{r}) &=& \myvec{\Phi}_{trap}(\myvec{r}) + \myvec{\Phi}_{inter}(\myvec{r})\\
\myvec{\Phi}_H(\myvec{r}) &=& \myvec{\Phi}_{trap}(\myvec{r}) - \myvec{\Phi}_{inter}(\myvec{r}) \\ 
\myvec{\Phi}(\myvec{r}) &=& 2 \myvec{\Phi}_{trap}(\myvec{r}),
\end{eqnarray}
with:
\begin{eqnarray}
\myvec{\Phi}_{trap}(\myvec{r}) &=& -\omega \frac{h_+ + h_-}{2} W_{trap}(\myvec{r}) \myvec{z}\\
\myvec{\Phi}_{inter}(\myvec{r}) &=& \omega W_{trap}(\myvec{r}) \bigg(h_1 cos\left(\delta\theta\right) cos\left(\varphi(\myvec{r})\right) + \notag \\
& & h_2 sin\left(\delta\theta\right) sin\left(\varphi(\myvec{r})\right)\bigg) \myvec{z} . \label{Appendix::Eq::Phi_EH}
\end{eqnarray}

Finally, we remind the expression for the chiral density $K(\myvec{r}) = \omega \varepsilon_f \mu_f {\rm Im}[\Esw\cdot\Hsw^*]/2 $, which corresponds to a simple trapping pattern modulated by the relative chirality of the beams 
\begin{eqnarray}
K(\myvec{r}) &=& - (h_+ - h_-)\cdot  \omega \sqrt{\varepsilon_f \mu_f} W_{trap}(\myvec{r}) . \label{Appendix::Eq::K}
\end{eqnarray}

These fluxes and potentials allow us to fully define the electric, magnetic and chiral forces 
\begin{eqnarray}
\myvec{F}_{E}^{reac}(\myvec{r}) &=& \Re[\alpha] \nabla W_E(\myvec{r})\notag \\
\myvec{F}_{H}^{reac}(\myvec{r}) &=& \Re[\beta]  \nabla W_H(\myvec{r})\notag \\
\myvec{F}_{\chi}^{reac}(\myvec{r}) &=& \Re[\chi] \frac{\nabla K(\myvec{r})}{\omega \sqrt{\varepsilon_f \mu_f}}\notag \\
\myvec{F}_{E}^{diss}(\myvec{r}) &=& \Im[\alpha] \left(\omega \varepsilon_f \mu_f \myvec{\Pi}(\myvec{r}) - \frac{\nabla\times\myvec{\Phi}_E(\myvec{r})}{\omega}\right)\notag \\
\myvec{F}_{H}^{diss}(\myvec{r}) &=& \Im[\beta] \left(\omega \varepsilon_f \mu_f \myvec{\Pi}(\myvec{r}) - \frac{\nabla\times\myvec{\Phi}_H(\myvec{r})}{\omega}\right)\notag \\
\myvec{F}_{\chi}^{diss}(\myvec{r}) &=& \sqrt{\varepsilon_f \mu_f} \Im[\chi] \bigg(2\myvec{\Phi}(\myvec{r})- \nabla\times\myvec{\Pi}(\myvec{r})\bigg) \label{Appendix::Eq::Forces}
\end{eqnarray}
that connect the real and imaginary parts of the electric-magnetic polarizabilities $\alpha$, $\beta$ and $\chi$ to the electric, magnetic and chiral densities and fluxes of the electromagnetic field.

In our configuration, the electric and magnetic dissipative forces are purely azimuthal, thus playing no role in the probability distributions of the double well. For our one-dimensional model, we can thus ignore them in the Fokker-Planck analysis where the only dissipative force that must be accounted for is the chiral dissipative force. Of course, these azimuthal components are accounted for in the three-dimensional simulations of the vectorial Langevin equation.

\section{Achiral force field landscape}
\label{Appendix:achiralLandscape}

\begin{figure*}[htb!]
  \centering{
    \includegraphics[width=\textwidth]{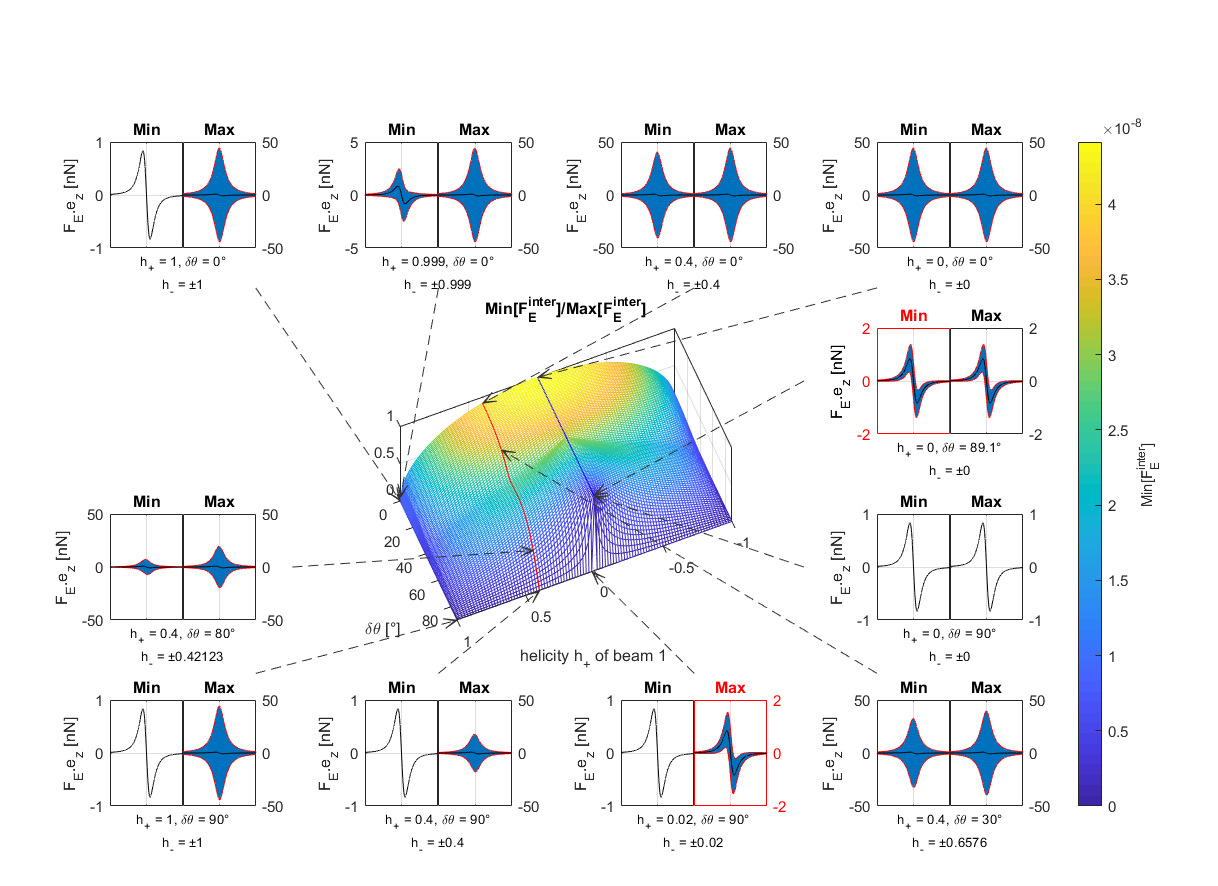}
    }
  \caption{General $(h_+,h_-,\delta,\delta\theta)$ polarization parameter space describing the achiral electric force field landscape whose equations are detailed in Appendix \ref{Appendix:Force_model} in terms of the ratio between the minimum and the maximum amplitude of the interference forces. All insets display forces along the optical axis  from both sides of the waist over a length of $\SI{20}{\micro\meter}$. Note that the force amplitudes of the insets vary from configurations to configurations.    }
  \label{Appendix:Fig:EHForceLandscape}
\end{figure*}

We here describe the general polarization parameter space in which the achiral electric force field landscape develops, as illustrated in Fig. \ref{Appendix:Fig:EHForceLandscape}. By choosing the $h_+$, $h_-$, and $\delta\theta$ parameters, we can tune the relative amplitude of the interference $\Re[\alpha]\grad{W_{inter}}$ (blue line of the insets) and average trapping forces $\Re[\alpha]\grad{W_{trap}}$ (black line of the insets of Fig. \ref{Appendix:Fig:EHForceLandscape}), while the $\delta$ parameters introduces a phase change in the interference forces. In Fig. \ref{Appendix:Fig:EHForceLandscape}, as well as in the rest of the paper, we systematically choose $\delta$  so that the interference potential is maximum at the center of the trap ($z = 0$). The surface represents the ratio between the maximum and the minimum amplitude of the interference forces when varying $h_-$ against the choice of $h_+$ and $\delta\theta$. 

The insets displayed in the figure showcase a few archetypical configurations. For a given choice of $h_+$ and $\delta\theta$, we present the choices of $h_-$ where the amplitude of the interference forces are respectively minimum and maximum. These two configurations are set with two opposite values of $h_-$ that are specified in the insets. The oscillations of the interference force in blue appear as a blue surface due to the high frequency of the oscillations with respect to the extension chosen for the optical axis. The red lines in the insets are the envelope of the total force. If we vary $\delta\theta$ along the $h_+ = 0$ line (in blue on the surface), the intensity of the interferences is varied, but does not change depending on $h_-$. The achiral landscape used in the main text is a more extreme case of the inset outlined in red on the right. It is shown in detail in Fig. \ref{Appendix:Fig:EHForceLandscapeZoom} below. The other red outlined inset is a reactive configuration --as the maximum of interference is obtained for opposite values oh $h_\pm$-- close to the one chosen in the main text. 

The dissipative configuration is an intermediate case where $\delta\theta \neq \SI{90}{\degree}$ allows for a purely dissipative interference force landscape. Along the red line on the surface, we vary $\delta\theta$ from $\SI{0}{\degree}$ to $\SI{90}{\degree}$, allowing for such intermediate cases to appear. Finally, in the configurations where $\delta\theta = \SI{90}{\degree}$ or $h_+ = 1$, we ensure that the minimum of interferences is always $\num{0}$ while for $\delta\theta = \SI{0}{\degree}$, we ensure that the maximum of interference is the global maximum. For other intermediate values of $\delta\theta$, the choice of $h_+$ can tune the minimum of interferences, ensuring that they are present, as seen for example following the red line.

\begin{figure}[htb!]
  \centering{
    \includegraphics[width=\linewidth]{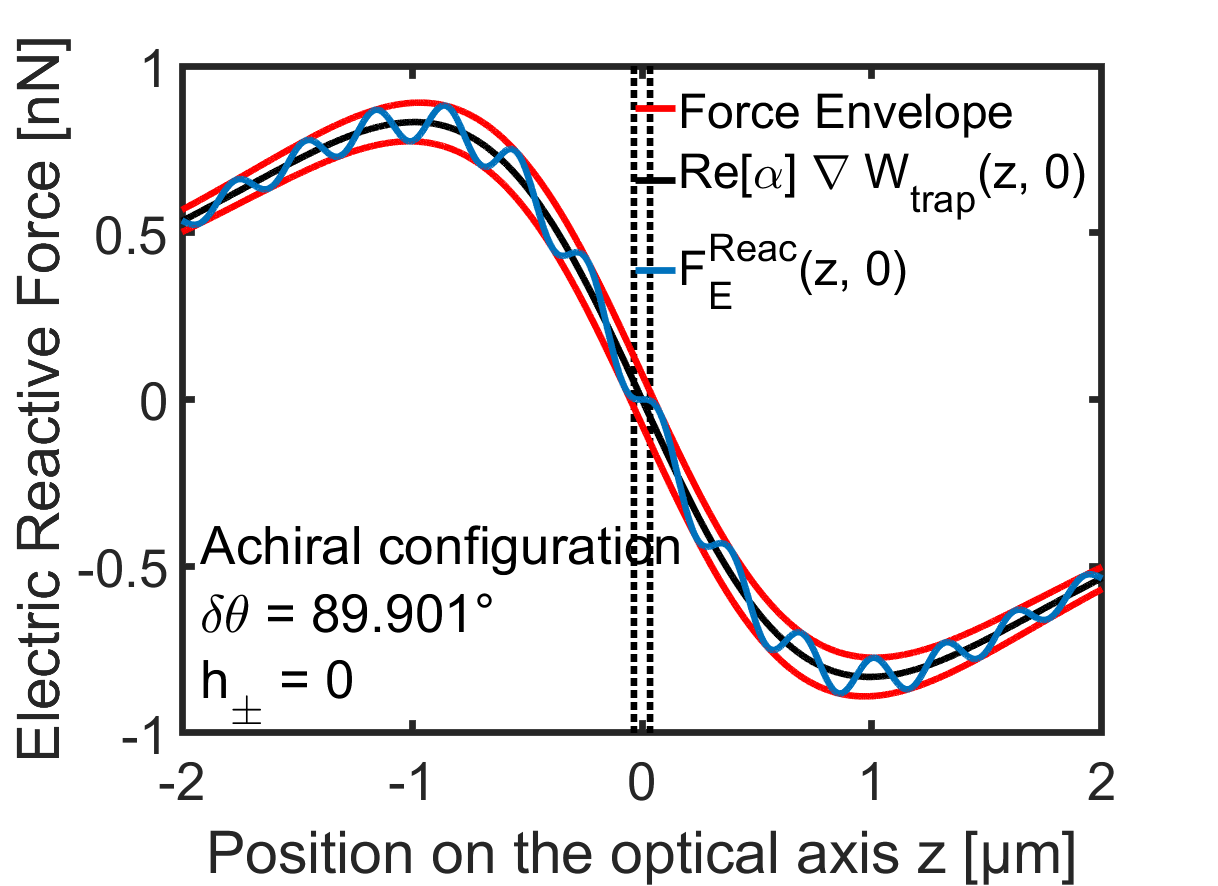}
    }
  \caption{Zoom on the electric force landscape in the achiral case, keeping in mind that the magnetic force is practically zero. This landscape is modified by the chiral forces when the helicity parameters $h_\pm$ are non zero. As in Fig. \ref{Appendix:Fig:EHForceLandscape}, the black line represents the force deriving from the trapping potential $-\Re[\alpha] \gradient{W_{trap}}(z, q)$. This force carries the contribution of the interference potential $-\Re[\alpha] \gradient{W_{inter}}(z, q)$ due to the interplay between the two counter-propagating beams. The total reactive electric force generated by these two components is represented by the blue line. Finally, the red line is the envelope of the electric forces. In the main text, we consider the nanospheres trapped at the center of the well at $z = 0$, with Fig. \ref{Fig:AchiralPotential} (b) displaying the force landscape evaluated between the dotted black lines. There is locally a double well caused by the oscillation due to the interference force around $z = 0$. In this configuration, we only have 2 wells, due to the fact that the interference forces are not strong enough to generate other trapping locations, as can be seen in the fact that they do not cross $\num{0}$ except at the center, and this despite their varying intensity.   }
  \label{Appendix:Fig:EHForceLandscapeZoom}
\end{figure}

\section{Dipolar chiral nanoparticle model}
\label{Appendix:polarizabilities}

Here, we follow \cite{canaguier2015chiral} in order to calculate the dipolar $(\myvec{p},\myvec{m})$ response of a chiral nanosphere that can be expressed in terms of the incident electric and magnetic fields $\myvec{E}$ and $\myvec{H}$
\begin{eqnarray}
\left( \begin{array}{c} \myvec{p} \\ \myvec{m} \end{array} \right) =  \left( \begin{array}{cc} \alpha \varepsilon_f &    i\chi \sqrt{\varepsilon_f \mu_f}   \\ -i\chi   \sqrt{\varepsilon_f / \mu_f} & \beta \end{array} \right)   \left( \begin{array}{c}\myvec{E} \\ \myvec{H} \end{array} \right).
\end{eqnarray}
In the quasistatic limit for a sphere of radius $R$, the electric, magnetic and chiral susceptibilities $\alpha$, $\beta$ and $\chi$ are given as:
\begin{eqnarray}
\alpha &=& 4 \pi R^3 \frac{(\epsilon_m - \epsilon_f)(\mu_m + 2\mu_f)-\kappa_m^2}{(\epsilon_m + 2\epsilon_f)(\mu_m + 2\mu_f)-\kappa_m^2} \\
\beta &=& 4 \pi R^3 \frac{(\epsilon_m + 2\epsilon_f)(\mu_m - \mu_f)-\kappa_m^2}{(\epsilon_m + 2\epsilon_f)(\mu_m + 2\mu_f)-\kappa_m^2} \\
\chi &=& 12 \pi R^3 \frac{\kappa_m}{(\epsilon_m + 2\epsilon_f)(\mu_m + 2\mu_f)-\kappa_m^2} , \label{eq:clausius}
\end{eqnarray}
where $\epsilon_m$ and $\mu_m$ are the complex permittivity and permeability of the material (in our case, gold), $\epsilon_f$ and $\mu_f$ are those of the fluid (assuming that both are purely real) and $\kappa_m$ is the complex ``chiral parameter'' of the nanosphere \cite{canaguier2015chiral}. 

In practice, for a non spherical chiral particle of arbitrary geometry, it is reasonable to assume that these equations will still apply with an effective electromagnetic radius $R$ and a chiral parameter that depends on the geometry of the particle. Exact equations can however always be calculated knowing the particular geometry. Experimentally, a determination of the complex chiral parameter can be obtained by measuring for the chiral nanoparticle the optical rotatory dispersion (ORD) for ${\rm Re}\left[\kappa_m\right]$ and the circular dichroism (CD) for ${\rm Im}\left[\kappa_m\right]$. 

From the chiral optical force perspectives, the polarizability $\chi$ is the relevant parameter, more precisely the ratio $\xi = \chi/\alpha$ that we fixed at a $5\%$ value throughout the Article. The Clausius-Mossotti relations (\ref{eq:clausius}) then lead to a simple relation that allows us to determine $\kappa_m$ from the chosen value for $\xi$:
\begin{eqnarray}
\kappa_m^2 + 3\kappa_m/\xi - (\epsilon_m - \epsilon_f)(\mu_m + 2\mu_f) = 0.
\end{eqnarray}

Among the four possible solution values for $\kappa_m$, we use the two opposite ones that have the smallest modulus. This choice is made in order to ensure that the transition from the achiral $\chi=0$ case to the chiral $\chi\neq 0$ case has practically no impact in the $\alpha$ and $\beta$ values. In such a case, the trapping potential profile described in Sec. \ref{sec:bistable} in the achiral case remains unchanged when the chiral coupling is induced in Secs. \ref{sec:reac} and \ref{sec:diss} with $\chi\neq 0$ and non-zero chiral density and/or flux.

\section{Simulations: algorithms and methods}
\label{Appendix:Simu}

The Langevin dynamics of an overdamped brownian object at position ${\bf r}$ immersed in a force field ${\bf F}$ and a fluid of viscosity $\gamma$ and diffusion coefficient $D$ is given by the equation 
\begin{eqnarray}
\dd{{\bf r}} &=& \frac{1}{\gamma}{\bf F}\dd{t} + \sqrt{2 D} \dd{{\bf W}_t}
\end{eqnarray}
where $\dd{{\bf W}_t}$ is the brownian increment at time $t$ 

To simulate the Langevin dynamics of a dipolar chiral particle in an axisymmetrical force field as done in Secs. \ref{sec:simuEns} and \ref{sec:simuSin}, we use the Euler-Maruyama scheme \cite{kloeden1992} 
\begin{eqnarray}
\rho_{n} &=& \sqrt{x_{n}^2+y_{n}^2}\notag\\
x_{n+1} &=& x_{n} + \frac{\dd{t}}{\gamma} \frac{x_{n} F_{\rho}(\rho_{n}, z_{n}) - y_{n} F_{\theta}(\rho_{n}, z_{n})}{\rho_{n}} \notag\\
        & & + \sqrt{2 D \dd{t}} \cdot \eta_{x}(n)\\
y_{n+1} &=& y_{n} + \frac{\dd{t}}{\gamma} \frac{y_{n} F_{\rho}(\rho_{n}, z_{n}) + x_{n} F_{\theta}(\rho_{n}, z_{n})}{\rho_{n}} \notag\\
        & & + \sqrt{2 D \dd{t}} \cdot \eta_{y}(n) \\
z_{n+1} &=& z_{n} + \frac{\dd{t}}{\gamma}  F_{z}(\rho_{n}, z_{n}) + \sqrt{2 D \dd{t}} \cdot \eta_{z}(n)
\end{eqnarray}
 where during the time increment $\dd{t}$, the Brownian increment on each axis is randomly chosen in the distribution $\eta_{x/y/z} = \sqrt{\dd{t}}\mathcal{N}(0,1)$. The simulation time step parameter $\dd{t}$ is chosen such that ${\rm Max}[F_i \sqrt{dt}/\sqrt{2k_{\rm B}T\gamma}]\leq 1$ for all cylindrical components $i=\rho,\theta,z$ of the optical force ${\bf F}$ (achiral and chiral).

In an effort to further reduce the calculation time and thus allow for better statistics to be used, we used the result of our one-dimensional model and draw the initial positions from the predicted stationary PDF in order to avoid the equilibration time. To do that, we use a multidimensional inverse transform sampling method. 

In a standard one-dimensional inverse transform sampling, knowing the distribution's PDF $p(X)$, we calculate the monotonic cumulative distribution function (CDF) $F(X)$. It can then be proved that if we draw a random number $U$ following a uniform distribution, $F^{-1}(U)$ will follow the distribution $p(X)$. In order to adapt this method to our multidimensional case, we first note that the problem being fully axisymmetrical, the azimuth $\theta$ can simply be chosen as a uniformly distributed random number. The two remaining parameters are then the axis and radius coordinates $z$ and $q$. 

As in the one-dimensional method, we calculate the PDF $p(q, z)$ obtained using our pseudo-potential model described in Section \ref{sec:diss} as $p(q,z)=C\exp \left[-\varphi(q,z)\right]$, with $\varphi(q,z)=(U_{\rm opt}(q,z)+u_{\chi}^{\rm diss}(q,z))/k_{\rm B}T$ and $C = \int\limits_0^{+\infty}\int\limits_{-\infty}^{+\infty}\exp \left[-\varphi(q,z)\right]\dd{z}2\pi q \dd{q}$. Its CDF $F(z, q)$ is defined by 
\begin{eqnarray}
F(z, q) &=& \int\limits_{-\infty}^{z}\int\limits_0^{q} p(q', z') 2\pi q' \dd{q'}\dd{z'}.
\end{eqnarray}
Since $p(q, z)$ has a complicated expression that cannot be easily inverted or integrated, we calculate $F(z, q)$ numerically over a large enough domain $[-z_M ; z_M]$ for $z$ and $[0 ; q_M]$ for $q$ and numerically perform the necessary inversions. We can then consider $F_z(z) = F(z, +\infty)$ and apply the inverse transform sampling method using $F_z(z)$ to pick a random number $z_c$ following the distribution $p_z(z) = \int\limits_0^{+\infty}p(q, z) 2\pi q\dd{q}$. In this context, it means that picking a random number $\eta_z$ in the uniform distribution on $[0 ; 1[$, we can find $z_c = F_z^{(-1)}(\eta_z)$. Finally, we can define $\eval{F_q}_{z=z_c}(q) = F(z_c, q)/F(z_c, +\infty)$ and use again the inverse transform sampling method to pick up a random number $q_c$ in the distribution $\eval{p_q}_{z = z_c}(q) = p(q, z = z_c)$. To do that, we again pick a random uniformly distributed number $\eta_q$ in $[0 ; 1[$ and apply $q_c = \eval{F_q}_{z=z_c}^{(-1)}(\eta_q)$. The pair $(q_c, z_c)$ of generated numbers thus follows the distribution $p(q, z)$. 

Repeating this method for each trajectory, we generate the initial distribution for our simulation using the stationary predictions from our one-dimensional model PDF. If this distribution were not the stationary distribution, it would relax towards it in the course the simulation, leading to a significant time spent in stabilizing the distribution rather than generating usable data. The one-dimensional model induces only errors small enough that the possible relaxation of the PDF parameters is dominated by their intrinsic thermal fluctuation. By generating a large number of steady-state trajectories, we can however check that using this distribution, the statistical parameters do not change in a measurable way over the simulated time. Therefore, all the generated time steps can be used for the data analysis of the properties of our simulated system in its steady-state. 

\section{Residence time probability density functions. }
\label{Appendix:Jump_PDF}

Sec. \ref{sec:simuSin} analyzes the distribution of the residency times in both wells A and C of the optical potential energy as a function of the presence and nature of the chiral coupling. These residence time are calculated using $\num{4096}$ long $\num{3000000}$ steps trajectories. We describe in this section how the residency times are identified. 

A diffusing trajectory in the bistable potential is characterized by different jump-like events. For some, the particle moves quickly from one well to the other. For many others however, the particles diffuses around the top of the unstable barrier or barely crosses it and returns back to its initial wells, so-called recrossing events. 

Following \cite{Schuetz2015}, we choose to use an hysteresis criterion to filter out such recrossing events. To do this, we exploit the repulsive character of the barrier which strength is given by a steepest descent approach similar to the one developed in Sec. \ref{sec:fokker} as $-\pdv[2]{\varphi_{eff}}{z}$ --where $\varphi_{eff}=U_{\rm opt}+U_{\rm \chi}^{\rm reac}+u_{\rm \chi}^{\rm diss}$ is evaluated on the optical axis $(q=0,z)$ including the chiral reactive potential $U_{\rm \chi}^{\rm reac}$ and/or dissipative pseudo-potential $u_{\rm \chi}^{\rm diss}$ depending on the chiral coupling cases.

This trapping strength leads to a standard deviation delimiting an exclusion zone of $\sigma = \sqrt{-k_B T / \pdv[2]{\varphi_{eff}}{z}} = \SI{10}{\nano\meter}$. We use this standard deviation to define the hysteresis of the bistability: the particle enters or leaves well A when it crosses the $z = -\sigma$ and enters or leaves well C when it crosses $z = \sigma$. But in addition, a jump is counted only when the opposite well is reached. In other words, a particle that would make an excursion in the vicinity of the barrier $z_B$ and eventually going back to its initial well will not be counted as having left its well. Such sequences are excluded from the record, as seen in black on Fig. \ref{Fig:Jumps} in the main text.

Having defined the crossing events, as shown in Fig. \ref{Fig:Jumps} (a) and (c), we measure the time interval $\tau$ that a particle has stayed in one well before jumping to the other. Because it is impossible to determine this time interval at the beginning and end of the trajectory, the corresponding events are excluded from the analysis. 

We then calculate the PDF of the occupation times of both wells. The results are show in Fig. \ref{Fig:Jumps} (b) and (d). According to Kramers theory, this PDF should follow an exponential law. However, we clearly observe deviations from such a law at short times, where the position of the particle remains correlated. The correlation time being $t_{\rm corr} = 2\pi\gamma/|\pdv[2]{\varphi_{eff}}{z} |_{(0,z_{A/C})}$, we therefore exclude from our analysis all traces recorded for times smaller that $t_{\rm corr}$. This being done, we finally perform a weighted fit of the distribution to take into account the fact that the smaller the probability, the lower the signal-over-noise ratio is. This fit yields precise values for the slopes of the exponential law --plotted in a logarithmic plot as represented Fig. \ref{Fig:Jumps} (b) and (d). From the Poissonian exponential law, these slopes correspond to the average residence time. 

\end{appendix}

\bibliography{biblio-chiral-thermo}

\end{document}